\documentclass[%
reprint,
amsmath,amssymb,aps,prb,dviepsmx 
]{revtex4-1}
\usepackage{graphicx}
\usepackage{dcolumn}
\usepackage{bm}
\usepackage{comment}
\usepackage{caption}
\usepackage{amsmath}
\usepackage{amssymb}
\usepackage{relsize}
\usepackage{braket}
\usepackage{ulem,times}
\usepackage{here}
\usepackage{color}
\usepackage{comment}
\usepackage{newtxtext,newtxmath}
\newcommand{\FRAC}[2]{{\textstyle \frac{#1}{#2}}}

\newcommand{\dd}{\bm{d}}
\newcommand{\nn}{\bm{n}}
\newcommand{\rr}{{\bm{r}}}
\newcommand{\qq}{{\bm{q}}}
\newcommand{\QQ}{\bm{Q}}
\newcommand{\jj}{\bm{j}}

\newcommand{\bphi}{\bm{\phi}}

\newcommand{\mcH}{\mathcal{H}}

\newcommand{\mcS}{\mathcal{S}}
\newcommand{\mcT}{\mathcal{T}}

\newcommand{\msm}[1]{\mspace{-#1 mu}}
\newcommand{\msp}[1]{\mspace{#1 mu}}

\newcommand{\behat}{\hat{\bm{e}}}
\newcommand{\bvhat}{\bm{v}} 

\newcommand{\vth}{\eta}
\newcommand{\vph}{\alpha}

\newcommand{\EQ}{\msm3 = \msm3}
\newcommand{\EQUIV}{\msm3 \equiv \msm3}
\newcommand{\GT}{\msm3 > \msm3}
\newcommand{\LT}{\msm3 < \msm3}

\newcommand{\CDOT}{\msm3 \cdot \msm3}

\newcommand{\PLUS}{\msm3 + \msm3}
\newcommand{\MINUS}{\msm3 - \msm3}

\usepackage{hyperref}
\usepackage{url}
\allowdisplaybreaks[4]

\begin{document}

\title{
Orbital moir\'e and quadrupolar triple-q physics in a triangular lattice
}
\author{Kazumasa Hattori}

\email{hattori@tmu.ac.jp}
\affiliation{%
Department of Physics, Tokyo Metropolitan University,
1-1, Minami-osawa, Hachioji, Tokyo 192-0397, Japan}
\author{Takayuki Ishitobi}
\affiliation{%
Advanced Science Research Center, Japan Atomic Energy Agency, Tokai, Ibaraki 319-1195, Japan}

\author{Hirokazu Tsunetsugu}
\affiliation{%
The Institute for Solid State Physics, 
The University of Tokyo, 
Kashiwanoha 5-1-5, Chiba 277-8581, Japan }

\date{\today}

\begin{abstract}
We numerically study orders of planer type $(xy,x^2-y^2)$ quadrupoles  
on a triangular lattice with 
nearest-neighbor isotropic $J$ and anisotropic $K$ interactions. 
This type of quadrupoles possesses unique single-ion anisotropy 
proportional to a third order of the quadrupole moments. 
This provides an unconventional mechanism of triple-$\qq$ orders which does not exist for the degrees of freedom with 
odd parity under time-reversal operation such as magnetic dipoles.  
In addition to several single-$\qq$ orders, 
{we find} various orders including incommensurate triple-$\qq$ quasi-long-range orders {with orbital moir\'e} and a four-sublattice triple-$\qq$ 
partial order. 
Our Monte-Carlo simulations demonstrate that the phase transition 
to the latter triple-$\qq$ state belongs to the universality class of 
the critical line of the Ashkin-Teller model in two dimensions close to the four-state Potts class. 
These results indicate a possibility of realizing unique quadrupole textures in simple triangular systems.
\end{abstract}

\pacs{Valid PAY'S appear here}
\maketitle


\textit{Introduction.---}
Superposition of multiple standing waves 
creates a moir\'e texture, and this is recognized in our daily life, e.g., when looking at overlapped fences, lace curtains swaying in the wind, and known as notorious wavy patterns in photography and its post processing \cite{amidror2009theory}. 
This visually fascinating texture is characterized by multiple wave vectors $\{{\qq}_\ell \}$.
Recent evolution of ``moir\'e engineering'' has been developed \cite{Xiao2020-cw,Kennes2021-pl} 
for twisted graphene multilayers \cite{Yankowitz2019-fi,Chen2019-sp,Wang2024-jj} 
and transition metal 
dichalcogenides \cite{Li2014-ve,Hippalgaonkar2017-cj,Lin2018-qy} and this 
opens a novel way to manipulate their functionality 
and physics itself such as superconductivity \cite{Yankowitz2019-fi}, 
nonreciprocal phenomena \cite{Lin2022-pz}, flat bands \cite{Al_Ezzi2024-nm} 
and magnetoelectric effects \cite{He2020-lz,Xie2022-jm}. 

Another realization of
moir\'e structure uses topological texture of internal degrees of freedom. Representative examples are superfluid He \cite{Mineev2013}, various types of superconductors \cite{Ivanov2001-pb,Garaud2012-ur,Zyuzin2017-zz}, Bose-Einstein condensates 
in cold atoms \cite{Abo-Shaeer2001-qw,Takeuchi2021-hh,Chen2020-uw}, and exciton-polariton condensates \cite{Rubo2007-ke,Lagoudakis2009-gn,Flayac2010-lh}. This started in magnetic systems from the seminal work about quasi long-range orders in planer magnets \cite{KT1973} and this has opened a way toward topological physics. The discovery of skyrmion crystal 
in MnSi \cite{Rossler2006-cq,Muhlbauer2009-bh,Tonomura2012-rm} triggered current explosion of researches which aim  realizing  moir\'e structure \cite{Tokura2021-ts}.  
Such a spin moir\'e \cite{Shimizu2021-qo,Shimizu2022-aq} is 
also realized by a crystallization of various topological defects \cite{Kanazawa2012-gg,Ishiwata2020-gh,Gobel2021-yq}. Clarifying their properties is important for understanding the fundamental physics and also for developing the next generation technology.   

Spatial modulations of a moir\'e structure are 
characterized by a set of ordering vectors
$\{ \qq _\ell \}$ ($\ell=1,2,\cdots$). 
This requires a multiple-$\qq$ order of order parameter, and   
control of  the ordering vectors $\{ \qq _\ell\}$ is an important issue. 
Ferromagnets with noncentrosymmetric structure have  
the Dzyaloshinskii-Moriya interaction \cite{Fert2023-dg} 
and it shifts the ordering vector $\qq$ 
from $\bm{0}$ to finite incommensurate (IC) values.  
Such an IC ordering is also established by other mechanisms such as  
geometrical frustration \cite{Okubo2012}, Fermi-surface effects in 
 itinerant electron systems \cite{Akagi2012-cc,Hayami2021-jl}, 
and anisotropic interactions \cite{Wang2021-yr,Yambe2022-qk,Yambe2023-kg}. 
These varieties of possibility opens a wider range of material research 
for realizing multiple-$\qq$ orders \cite{Hirschberger2021-iu}.  

Recently, several studies have tried  realizing  
multiple-$\qq$ orders of nonmagnetic degrees of freedom $\varphi$ such as electric charge \cite{Ortiz2019,Denner2021} 
and electric quadrupole \cite{Ishitobi2021,Yanagisawa2021,Ishitobi2023UNi4B}. 
These order parameters are time-reversal even, which differs from magnetic ones. This 
makes a crucial difference in their free energy 
from the magnetic counterpart 
\cite{Walker_1994,Nikolaev1999-yt,Denner2021,Tsunetsugu2021,Hattori2023-pj}, and, for example, the cubic terms 
$\varphi_{\alpha} \varphi_{\beta} \varphi_{\gamma}$ are allowed. 
Such terms may arise from a single-ion 
anisotropy and favor triple-$\qq$ orders, 
since the term  
$\propto \varphi_{\alpha} (\qq_1) \varphi_{\beta} (\qq_2) 
 \varphi_{\gamma} (\qq_3)\delta_{\qq_1+\qq_2+\qq_3,\bf 0}$ lowers the free energy by properly choosing the phase and direction of each $\varphi(\qq_\ell) $.  { In magnetic systems, such couplings are possible only when a finite magnetic field is applied \cite{Wang2021-yr}. } Thus, nonmagnetic systems have a potential to exhibit multiple-$\qq$ physics { at zero magnetic field}, 
and provide a good playground for 
the moir\'e engineering. 

In this Letter, 
we propose a minimal model and 
 demonstrate that  nonmagnetic systems are useful for realizing 
moir\'e textures. The model is defined for electric quadrupoles on a triangular lattice 
with nearest-neighbor interactions and 
we have performed Monte Carlo simulations for its effective classical Hamiltonian. 
In addition to varieties of phases, we have found IC triple-$\qq$ quasi long-range orders relevant to the
two-dimensional Ashkin-Teller and Potts models with the partial
ordering of quadrupole moments.

\textit{Model.---}We consider a triangular lattice with the hexagonal site symmetry of the point group
D$_{6}$ or C$_{6v}$. 
In this case, the order parameter with the symmetry of 
$\Gamma_5$ (E$_2$) doublet is
particularly interesting as will be explained below. 
This is the two-component quadrupole 
$(\phi_u , \phi_v )$ with the symmetry of $(2xy, x^2-y^2)$, 
and we ignore magnetic sectors, which might be present in real materials, for simplicity. We use a classical approximation to represent this quadrupole moment 
by a unit vector pointing to the 
 direction denoted by $\theta_{\rr}$
\begin{equation}
\bphi _{\rr} = ( \phi_{u, \rr}, \phi_{v,\rr} ) =:  
\behat ( \theta_{\rr}) , \ \ \ 
\behat (\theta) \equiv ( \cos \theta, \sin \theta ),  
\label{eq:phi_r}
\end{equation}
where $\rr$ is the site position. 
We will use $\behat$ also for other vectors 
along with $\behat _{\ell}^{} \EQUIV \behat (\ell \msp1 \omega )$
for the special angle $\omega \equiv \FRAC{2\pi}{3}$. This classical description is valid when its finite temperature properties are discussed. 
These two-component quadrupoles 
possess a local cubic coupling as a single-ion anisotropy: 
\begin{equation}
V= \lambda \sum_{\rr} 
( 3 \msp1 \phi_{u,\rr}^2 - \phi_{v,\rr}^2 ) \, \phi_{v,\rr}^{} 
= \lambda \sum_{\rr} \sin  3 \msp1 \theta_{\rr} . 
\label{eq:sin3theta}	
\end{equation}
Without loss of generality we can set $\lambda\ge  0$. Its microscopic origin is discussed in the Supplemental Material \cite{SM}  
for models with the doublet local ground state E$_2$: $\{2xy,x^2-y^2\}$ and a singlet excited state A$_1$: $3z^2-r^2$.

Now let us introduce exchange couplings for the quadrupole moments.  
Each nearest-neighbor pair $\bphi _{\rr}$ and $\bphi _{\rr+\dd}$ 
interact via two types of exchange couplings: 
an isotropic $J$ and an anisotropic $K$ one  as 
\begin{equation}
\mcH _{\rm ex} =\sum_{\rr,\dd} 
\sum_{ij}
\bigl[ 
 (J-K) \msp2 \delta_{ij} + 2 K \msp2 d_i d_j \bigr] 
\bphi _{i,\rr+\dd} \, \bphi _{j, \rr} . 
\label{eq:3}	
\end{equation}
Here, 
$\rr$ runs over the sites on the triangular lattice. 
$\dd$ runs over a half of the nearest neighbors 
$\dd\equiv(d_u,d_v) \msm2 \in \msm2 
\{ \behat _n^{} \}_{n=0}^2$.  
Then, the Hamiltonian is the sum of the two terms 
$\mcH \EQ \mcH _{\mathrm{ex}} \PLUS V$. Note that 
the orbital degrees of freedom are nothing but anisotropic charge distribution, and thus, they usually possess anisotropy in their  Hamiltonian. See Refs.~\cite{Vernay2004-fk,Khaliullin2021-jn,Changle-Liu-Yao-Dong-Li-and-Gang-Chen2018-ww}, for example. 
This contrasts to the magnetic systems where the single site anisotropy originates from small spin-orbit couplings.

{\it Ordering vectors.---}Let us first examine the single-$\qq$ spiral orders 
$\bphi _{\rr} \propto \mathrm{Re} \, \bphi (\qq^\star ) 
 \exp(\pm i\qq ^\star \CDOT \rr)$, and determine the value 
of ordering vector $\qq^\star$ \cite{SM}.  
One can easily find that the order is ferroic ($\qq^\star = \bm{0}$) 
when 
$2J<-|K|$. When 
$-|K|<2J<5|K|$, the order is commensurate antiferroic (AF) 
and $\qq^\star$ locates at one of the edge centers (the M points) in the Brillouin zone (BZ): 
$\QQ _{\ell}^{} \EQ \frac{2\pi}{\sqrt{3}} \, 
\behat ( ( \frac14 -\ell) \msp2 \omega )$ ($\ell \EQ 1,2,3$).  
An interesting part is the IC state for the region of 
$5|K|<2J$. 
 As $|K|$ decreases, $\qq^\star$ leaves $\QQ _{\ell}^{}$ 
and moves along the BZ edge:
$\qq_{\ell}^\star \EQ \QQ _{\ell}^{} \PLUS 
 \Delta_{\rm M} \, \behat_{4-\ell}$ with 
$\cos (\Delta_{\rm M}/2) \EQ (2J+|K|)/(4J-4|K|)$ 
and it reaches the BZ corner 
[K or K$'$ point; 
$\qq _{\mathrm{K}} \EQ 2\omega \behat _0$ 
and 
$\qq _{\mathrm{K}'} \EQ 2\omega \hat{\bm{e}}(\tfrac{5}{2}\omega)$] when 
$K  \EQ 0$. 
As shown in Figs.~\ref{fig:1}(a) and (b), the lowest-energy  eigenmodes $\bphi (\qq_{\ell}^\star)$ differ 
between distinct edges, 
but remain unchanged on each edge 
irrespective of $\Delta_{\rm M}$: 
$\bphi (\qq_\ell^\star) 
\propto 
\bvhat _\ell^{+} \EQ \behat( (\ell \MINUS \FRAC14) \msp1 \omega)$ 
or 
$\bvhat _\ell^{-} \EQ \behat _{\ell \MINUS 1}^{}$.


\begin{figure}[t!]
\includegraphics[width=0.5\textwidth]{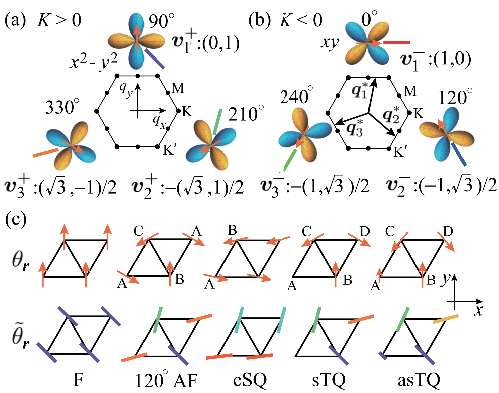}
\caption{
Lowest-energy eigenmode $\bm{\phi}(\qq^\star)$ of $\mcH _{\rm ex}$  for $\qq^\star_\ell$ on the BZ edges for 
(a) $K>0$ and (b) $K<0$.  
Within each edge, the eigenmode does not change 
 and the quadrupole moment is illustrated
together with the corresponding pseudo spin 
(arrow for angle $\theta$) and 
nematic director (bar for angle $\tilde{\theta}=-\frac{1}{2}\theta$).\hspace{2cm}    
(c) Representative configurations of $\theta_\rr$ and $\tilde{\theta}_{\rr}$ in commensurate orders; ferroic (F), 120$^\circ$ antiferroic (120$^\circ$ AF), canted single-$\qq$ (cSQ), symmetric triple-$\qq$ (sTQ), and asymmetric triple-$\qq$ (asTQ).  
A--D are the sublattice indices.}
\label{fig:1}
\end{figure}

\textit{Ferroic, AF 120$^\circ$, and triple-$\qq$ orders.---}Now 
that $\qq_{\ell}^\star$'s are determined,  the classical solution 
of $\bphi _{\rr}$ is readily obtained.  
However, the cubic potential (\ref{eq:sin3theta}) 
causes nontrivial modifications.  
Before their discussion, let us analyze simpler parts, \textit{i.e.}, 
the ferroic (F) order for $2J \LT -|K|$ and 
the 120$^\circ$ AF order for $5|K| \LT 2J$.  
See Fig.~\ref{fig:1}(c).  
Owing to the cubic term (\ref{eq:sin3theta}), 
these phases have no continuous degeneracy, 
and the moment's direction is locked at 
$\theta_{\rr} \EQ ( n_{\rr} \PLUS \FRAC34 ) \msp1 \omega$ 
with $n_\rr \msm2 \in \msm2 \{ 0,1,2 \}$. 

For  $0 \LT 2J \LT 5K$, the simplest candidate is the symmetric triple-$\qq$ (sTQ) state.  
The three ordering vectors locate at 
$\QQ _{\ell}^{}$'s with an identical amplitude 
$\phi_\ell\equiv | \bphi (\QQ _{\ell}^{}) |$ for all $\ell$.  
Note that $\bphi (\QQ _{\ell}^{})$'s are all real 
since $-\QQ _{\ell}^{} \EQUIV \QQ _{\ell}^{}$.  
The $\bvhat_{\ell}^{+}$ mode  has a lower energy of $\mcH _{\mathrm{ex}}$ 
than the $\bvhat_{\ell}^{-}$ mode's, 
and this comprises the triple-$\qq$ configuration
\begin{align}
\bphi _{\rr}=  \msm6 \sum_{\ell=1,2,3} \msm6 \phi_{\ell} \msp2 
\cos \bigl( \QQ _{\ell}^{} \msm2 \cdot \msm2 \rr) \, \sigma_{\ell} \, 
\bvhat_{\ell}^+, 
\hspace{3mm} 
\sigma _{\ell} \in \{ +1, -1 \} . 
\label{eq:tripleM}
\end{align}
Substituting this with 
$\phi_{\ell} \EQ \FRAC12 \phi \GT 0$ in Eq.~(\ref{eq:sin3theta}), 
one finds 
$V \EQ - \FRAC34 N\msp2 \lambda \msp2 \phi^3 
 \sigma _1 \msp2 \sigma _2 \msp2 \sigma _3$, where $N$ is the number of sites.  
This term is minimized when all thee or otherwise one of 
$\sigma_{\ell}$'s is $+1$. 
This energy gain readily guarantees that 
the first-order transition occurs at a temperature higher 
than that of a continuous transition to the single-$\qq$ state in the simple Landau analysis.  
This sTQ state indeed has a four-sublattice \textit{partial order} 
with $\{\bphi _{\rr}\} \EQ \{\bm{0}$,  
$\phi \msp2 \bvhat _1^+$, 
$\phi \msp2 \bvhat _2^+$, 
$\phi \msp2 \bvhat _3^+\}$ as shown in Fig.~\ref{fig:1}(c). 
The four sets of $\{\sigma _\ell \}$ correspond to 
domains with the  disordered site on different sublattices. 

For $ 5K \LT 2J \LT 0$, one may expect a similar sTQ state, 
but the situation is quite different. 
Employing Eq.~(\ref{eq:tripleM}) with $\bvhat _\ell^{+}\to \bvhat _\ell^{-}$  
results in  $V=0$. 
Thus, the sTQ state is not stable for $K<0$  in contrast to the case of $K>0$.  

\textit{Parasitic ferro moments.---}The cubic term  (\ref{eq:sin3theta}) affects 
the order parameters significantly. 
Using the eigenmodes 
$\bvhat _{\ell}^{+}$ for $K>0$, we find that $V$ contains the
following term
\vspace{-4pt}
\begin{equation}
-\FRAC32 \lambda \bm{D} \cdot \bphi (\bm{0}) , 
\hspace{3mm} 
\bm{D} \EQ 
\bigl( 
  \sqrt{3} ( \phi_3^2 \MINUS \phi_2^2 ), \,  
   2\phi_1^2 \MINUS \phi_2^2 \MINUS \phi_3^2 
\bigr).  
\end{equation}
Once the symmetry $\phi_1$=$\phi_2$=$\phi_3$ is 
 broken, $\bm{D}$ is activated and this induces the ferro moment $\bphi (\bm{0})$\cite{Tsunetsugu2021,Hattori2023-pj}.  
This is natural since in the sTQ state 
$\bm{\phi}_\rr$'s are disordered at a quarter of the sites and thus this state is not stable at $T=0$. 
Such an asymmetric triple-$\qq$ (asTQ) state is illustrated in Fig.~\ref{fig:1}(c) for the case of $\phi_1 \GT \phi_2 \EQ \phi_3$, and 
$\bphi (\bm{0}) \msm2 \parallel \msm2 \behat (\FRAC{\pi}{2})$. 
In general, the $\bvhat _{\ell}^{+}$ and $\bvhat _{\ell}^{-}$ components hybridize, 
and this rotates $\bphi _{\rr}$. 
See $\bphi_\rr$ in the C and D sublattices in Fig.~\ref{fig:1}(c).

Even for single-$\qq$ states, such a ferroic moment emerges.
Consider the single-$\qq$ state for $K \LT 0$ with 
$\qq^\star \EQ \QQ _{\ell}^{}$.  
The AF mode 
$\bphi ( \QQ _{\ell}^{} ) \EQUIV \phi_{\ell} \msp2 
\behat (\theta_{\ell})$ 
can couple with the ferro mode 
$\bphi (\bm{0}) \EQ 
\phi_0 \msp2 \behat (\theta_0)$, and this yields 
$ V\sim 
3N\msp1 \lambda \msp2 \phi_\ell^2 \msp2 \phi_0 \msp2 
\sin ( 2\theta_\ell \PLUS \theta_0)$.  
This shows that the ferro component $\bphi (\bm{0})$ 
points to the direction 
$\theta_0 \EQ \FRAC32 \pi \MINUS 2\theta_\ell$, 
and thus, the canted single-$\qq$ (cSQ) state is stabilized as shown 
in Fig.~\ref{fig:1}(c). 
For $K<0$, 
$\theta_{\ell} \EQ (\ell \MINUS 1) \msp1 \omega$, 
and thus,  
$\bphi (\bm{0}) \msm2 \parallel \msm2 
 \behat ((-\FRAC74 \PLUS \ell )\msp1 \omega)$.

\begin{figure}[t!]
\includegraphics[width=0.5\textwidth]{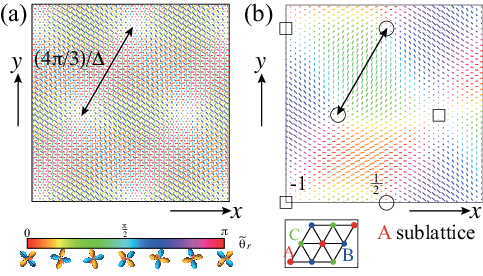}
\caption{
(a) Director angle configuration $\tilde{\theta}_{\rr}$ 
in the triple-$\qq$ state 
with 
$\qq_{\ell}^\star \msm2 \sim \msm2 
 \qq_{\mathrm{K}'} \MINUS \Delta \msp1 \behat _{4-\ell}$ 
with 
$\Delta \EQ \frac{\pi}{24}$. \hspace{2cm}
(b) 
Half-vortices ($\circ$) form a honeycomb lattice, and vortices ($\square$) are positioned at the center of each hexagonal unit. 
} 
\label{fig:2}
\end{figure}

\textit{IC triple-$\qq$.---}Now, we study the 
triple-$\qq$ states  for $K \GT 0$ with the ordering vectors 
$\{ \qq^\star_{\ell} \}_{\ell=1}^3$ 
on the BZ edges. 
The cubic term is important, as it stabilizes the IC triple-$\qq$ 
(IC-TQ) states. 
Since 
$-\qq_{\ell}^\star \msm2 \not\equiv \msm2 \qq_{\ell}^\star$, 
the factor 
$\sigma_{\ell} \msp1 \cos (\QQ _{\ell}\!\cdot\!\rr)$ in 
Eq.~(\ref{eq:tripleM}) should be replaced by 
$\cos(\qq_{\ell}^\star\!\cdot\!\rr \PLUS \vth _{\ell})$ 
with a general phase $\vth _{\ell}$.  
Then,  
$V \EQ -\FRAC34 N \lambda \msp2 \phi_1 \phi_2 \phi_3 \msp2 
\cos ( \vth _1 \PLUS \vth _2 \PLUS \vth _3)$
and this is minimized when 
$ \vth _1 \PLUS \vth _2 \PLUS \vth _3 \EQUIV 0$.  
This energy gain guarantees that the IC-TQ state is more stable than the 
single-$\qq$ state for sufficiently large $\lambda$. 

Such triple-$\qq$ configurations can be visualized as 
\textit{orbital moir\'e}. 
Figure \ref{fig:2}(a) shows an example of ``nematic'' 
director angles 
$\tilde{\theta}_{\rr} \EQUIV -\FRAC12 \theta_{\rr}$ 
(mod $\pi$).   They point to the direction of the orbital 
node: $\tilde{\theta}_{\rr} \EQ 0$ 
for the $xy$-orbital and
$\FRAC34 \pi$ for $x^2 \MINUS y^2$.  
When $\qq_{\ell}^\star$'s are near the K or K$^\prime$ 
point, the configuration is characterized by a set of 
sublattice directors $\tilde{\theta}_{\rr\in {\rm X}}$ 
(X=A,B, or C). 
Then, we introduce their vector vortex charges 
$\nn \EQ (n_{\mathrm{A}},n_{\mathrm{B}},n_{\mathrm{C}})$ 
defined by \cite{Ueda2016-pj}  
\begin{equation}
n_{\rm X} (\rr) 
\equiv 
\oint_{\rr + \rr' \in {\rm X}} 
\frac{d\rr '}{2\pi} \, \cdot  
\frac{\partial}{\partial \rr '} \msp2 
\tilde{\theta}_{ \rr + \rr '} , \ \ \ 
\end{equation}
where the integral is defined for a closed loop 
encircling the site $\rr$.  For the data in Fig.~\ref{fig:2}, 
the ordering vectors are chosen as 
$\qq_{\ell}^\star \EQ
 \qq_{\mathrm{K}'} \MINUS \Delta \msp1 \behat _{4-\ell}$ with
$\Delta \EQ \frac{\pi}{24}$.  
The $\tilde{\theta}_\rr$ configuration in the panel (a) shows a triangular lattice of vortices 
and half-vortices with the approximate periodicity 
$4\pi/(3\Delta)$. Their   vector vortex charges are 
either  $\nn _1 \EQ (-1, \FRAC12 , \FRAC12 )$, 
$\nn _2 \EQ (\FRAC12 , -1, \FRAC12 )$, or
$\nn _3 \EQ (\FRAC12 , \FRAC12 ,-1)$. 
The panel (b) shows the director configuration in the A-sublattice. One can easily see that half-vortices
$(n_{\mathrm{A}} \EQ \FRAC12)$ form a honeycomb structure with vortices $(n_{\rm A}$$=$$-1)$ being located at the center of each hexagon.  
 Similarly, when $\qq_\ell$'s are close to the M points, the moir\'e texture 
 is characterized by four-sublattice vortices \cite{SM}. 
These moir\'e textures have cores where $|\bm{\phi}_{\rr}|\simeq 0$. 
 These cores are energetically unfavored at low temperatures, similar to the disordered sites in the sTQ state. They are common characteristics of the triple-$\qq$ states at finite temperatures \cite{Okubo2012}. 

\textit{Monte Carlo simulation.---}We have 
performed classical Monte Carlo (MC) simulations 
to examine the effects of thermal fluctuations on the 
triple-$\qq$ orders discussed above. 
We have used hexagonal clusters with the edge length 
$L\le 96$ and the maximum one has $N\!=$27937 sites. 
 The periodic boundary conditions are applied for 
 the three pairs of 
 opposite edges. We use exchange MC 
 algorithm \cite{Swendsen1986-mm,Hukushima1996-xo} 
 combined with single-``spin'' flip updates for
efficient simulations. 
At each temperature $T$, we calculate 
observables by averaging over $M \EQ 10^6$ MC steps 
(MCS) after thermalization of typically $10^5$ MCS. 
We parameterize 
$(J,K) \EQ \bar{J} (\cos\vph ,\sin\vph )$ 
with $\bar{J}=1$.

Let $\{ \bphi ^{(m)} _{\rr} \}$ 
denote the snapshot at the $m$-th MCS and 
define their Fourier component by 
$
\phi_{\sigma}^{(m)} (\qq_{\ell}) \equiv N^{-1} 
\sum_{\rr}
\bvhat _{\ell}^{\sigma} \cdot \bphi _{\rr}^{(m)} \, 
e^{{\color{blue} -} i\qq_{\ell} \cdot \rr}$
$(\sigma \EQ \pm)$ 
for those $\qq _{\ell}$'s on the $\ell$-th BZ edge. 
Since 
$\phi_{\sigma}^{(m)} (\qq_{\ell})$'s may have  arbitrary phases, 
we define the single-$\qq$ order parameter as 
$\mcS ^\sigma ( \qq_{\ell} ) 
:= 
M^{-1} \sum_m 
\bigl| \phi _{\sigma}^{(m)} 
(\qq _{\ell}) \bigr|^2$ 
$=:$  
$\langle 
| \phi_{\sigma} (\qq_{\ell}) |^2
\rangle$, 
while that for the triple-$\qq$ order is defined as 
$\mcT ^{\sigma} (\{ \qq_{\ell} \}_{\ell=1}^3) \equiv 
\langle  
\phi_{\sigma} \msm1 (\qq_1) \msp2
\phi_{\sigma} \msm1 (\qq_2) \msp2
\phi_{\sigma} \msm1 (\qq_3) 
\rangle$
with $\qq_1 \PLUS \qq_2 \PLUS \qq_3 \EQUIV \bm{0}$. 

\begin{figure}[t!]
\includegraphics[width=0.5\textwidth]{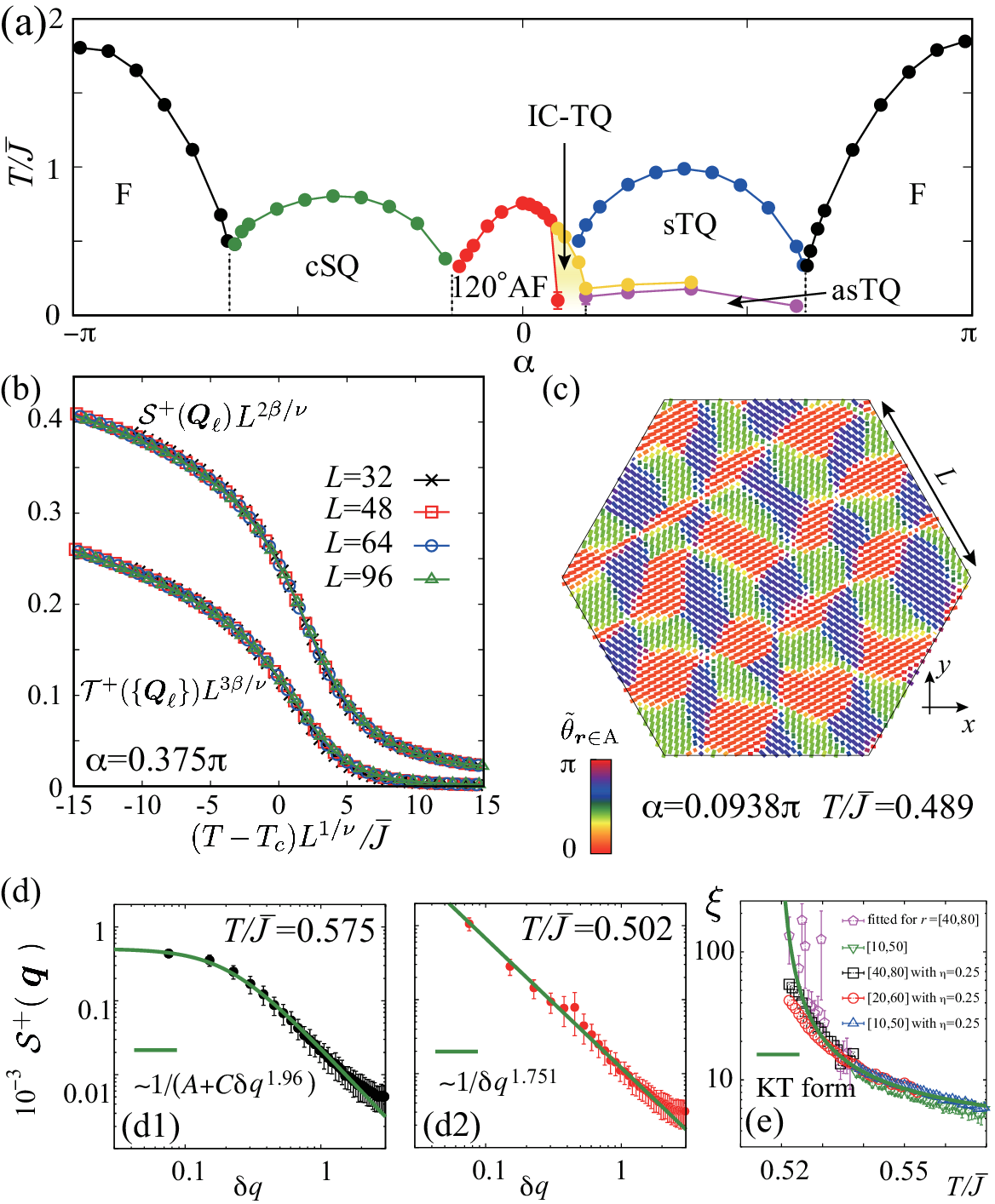}
\caption{
(a) $T$--$\vph $ phase diagram. 
The low-$T$ region of the IC-TQ state, 
situated  between the 120$^\circ$ and the sTQ 
states, is expected to be the 120$^\circ$-AF one,  
because $\qq^\star$ moves to $\qq_{\rm K,K'}$ as $T$ 
decreases.
(b) Finite-size scaling analysis 
of ${\mathcal S}^+(\bm{Q}_\ell)$ and 
${\mathcal T}^+(\{\bm{Q}_\ell\})$  for
$\vph =0.375\pi$ with the choice $(\nu,\beta)=(0.71 , 0.089)$, 
where the transition temperature is assumed 
$T_c/\bar{J}\simeq 0.9853$.  
(c) Snapshot $\tilde{\theta}_{\rr}$ in the 
A-sublattice for $L=50$, $\alpha=0.0938\pi$ and 
$T/\bar{J}$$=$$0.489$, where the IC-TQ configuration 
with $\qq_{\ell}^\star \msm2$$=$$\msm2 
 \qq_{\mathrm{K}'}$$\MINUS$ 
 $\tfrac{4\pi}{3}\tfrac{3}{50}\msp1 \behat _{4-\ell}$ 
 is realized. 
(d) $\mathcal S^+(\qq)$ vs $\delta \qq=\qq-\qq^\star$ 
for (d1) $T>T^*$ and (d2) $T<T^*$ for $L=96$, where 
the data along two directions $\delta\qq\parallel 
\behat(\ell\msp1 \omega)$ and 
$\parallel\behat((\ell \MINUS \FRAC14) \msp1 \omega)$ 
are averaged. The data are fitted in the region 
of $0.5\le |\delta\qq| \le 2.0$ by a function shown 
in the figure. 
(e) $T$ dependence of correlation length $\xi$ for 
$L=96$ obtained by fitting 
$\chi_{uu}(r\hat{\bm{e}}_0)$=
$c_1\cos[(c_2+\omega)r]$$r^{-c_3}e^{-r/\xi}$. The fitting range is  
$10\le r\le 50$ (blue, green) for higher $T$, while  
$20\le r\le 60$ (red) and $40\le r\le 80$ 
(black, pink) for lower $T$. The green curve 
is the KT form $\xi= a_1\exp[a_2/\sqrt{T-T^*}]$ 
with $a_{1,2}$ being a constant. Data fitted with 
fixed $c_3=1/4$ (expected in the KT form) and those 
with $c_3$ being a free parameter are similar, while 
the low-$T$ data for the latter have large error 
bars (pink).} 
\label{fig:3}
\end{figure}

Figure \ref{fig:3}(a) shows the $T$-$\vph$ phase 
diagram for $\lambda=1$. 
 Similar results have been obtained for $\lambda\gtrsim 0.5$ \cite{SM}.   
There appear six ordered phases in total.  
Their transition temperature is mostly determined as 
the temperature where the Binder ratio 
of the order parameter coincides for $L=32$ and $48$. 
For first-order transitions, it is identified 
to the temperature at which the order parameter jumps 
upon lowering $T$. 
Near $\vph \EQ 0$, 
the 120$^\circ$ AF order is stable and its transition 
from the high-$T$ disordered phase is first order. This is fully consistent with the results of the analysis based on the Landau theory \cite{SM}. The presence of six 120-degree AF domains suggests that the transition is described by the six-state Potts model, which is known to exhibit a first-order transition \cite{Wu1982-sm}. For the F and cSQ states, 
the finite-size scaling analysis shows that 
the transition is of the three-state Potts universality 
class \cite{SM}. 
For the sTQ, we find that 
the universality class of the transition belongs to that on the critical line of the Ashkin-Teller 
model \cite{Nienhuis}. This is apparent in both triple-$\qq$ 
 $\mcT ^{+} (\{ \QQ _{\ell}^{} \}_{\ell=1}^3)$  and 
 single-$\qq$ $\mcS ^+ (\QQ _{\ell}^{})$ order parameters 
 as shown in Fig.~\ref{fig:3}(b) for $\alpha=0.375\pi$. 
The critical exponents determined are 
$\nu \simeq 0.71$ and $\beta \EQ \nu/8$, where the 
latter relation holds along the critical line. 
As $\alpha$ varies, $\nu$ changes slightly, but its 
value is close to that for the four-state Potts  
universality class corresponding to the end point 
of the critical line: $\nu \EQ \FRAC23$ and 
$\beta \EQ \FRAC{1}{12}$ \cite{Wu1982-sm,Creswick_1997}.  
 These results are related to the 
four-fold degeneracy of the disordered site position. 
Similar situation has been discussed in the literature 
for such an example of N$_2$ molecules adsorbed on 
Kr-plated graphite \cite{Zia1975-no,Domany1977-yu}. 
Between the sTQ and 120$^\circ$-AF states, 
the IC-TQ appears at finite $T$. 
However, it becomes unstable at lower $T$'s since 
 a finite density of disordered sites
 has no contribution to energy gain. 
Note that the ``incommensurate'' is just nominal and 
means $\qq^\star$ is neither on the K(K$^\prime$) nor M point along the edge of the BZ.

\textit{IC-TQ and orbital moir\'e.---} In the IC-TQ state, 
the ordering vector $\bm{q}^\star$ moves  
along the BZ edge with lowering $T$, and 
this is similar to the case of the Devil's staircase 
in frustrated magnets \cite{Bak1980-ez}. 
The varying $\qq^\star$ is triggered by forming 
dislocations of the vortex lattice \cite{SM}.
Figure \ref{fig:3}(c) shows a snapshot 
sublattice configuration of $\tilde{\theta}_\rr$ 
averaged over 1000 MCS. It is well described by a 
hexagonal moir\'e texture with its sublattice vortices 
 $n_{\mathrm{A}} \EQ -1$ and $\FRAC12$ (Fig.~\ref{fig:2}). 
 The short-time averaging makes the vortex cores clearer 
 than in a single snapshot. For 
 $\alpha\lesssim \rm{tan}^{-1}\tfrac{2}{5}\simeq \alpha_c$, 
 the 120$^\circ$ AF order seems to appear at the lowest 
  temperature. Near $\alpha=\alpha_c$, the MC sampling is 
  not sufficient due to the presence of many local minima
  related to the IC configurations. We expect $\qq^\star$ 
approach $\qq_{\rm K,K'}$ and the IC-TQ states are 
  unstable at lower temperatures. These IC-TQ states 
intervene between the sTQ and the asTQ states for 
$\alpha\gtrsim \alpha_c$. However, we should be 
aware that the vector $\qq^\star$ depends on $L$ and 
the boundary conditions for these IC-TQ states. 
 Recalling that the phases $\eta_{\ell}$'s correspond 
 to the choice of the origin of IC oscillation 
 \cite{Shimizu2021-qo}, it is natural to expect 
 their fluctuations destroy the IC long range 
 order \cite{KT1973, Nelson1979-af,Coppersmith1982-im,Agterberg2011-cn}. 
 Indeed, one evidence of such a quasi-long-range 
 order is the power-law behavior of the correlation 
 function of $\bphi(\qq)$  shown in Fig.~\ref{fig:3}(d), 
 where the mass term $A$ in 
 $\mathcal S^+(\qq)$$\sim$
 $(A+C|\qq-\qq^\star|^2)^{-1}$ vanishes below 
 the transition temperature $T^*/\bar{J}\simeq 0.518$. 
 The dependence on $\delta\qq=\qq-\qq^\star$ is obtained 
 by fitting the data for $0.5<|\delta \qq|<2.0$ near 
 $T^*$, resulting in 
 $\mathcal S^+(\qq)\sim |\delta\qq|^{-\zeta}$ with 
 $\zeta\simeq$1.6$\sim$2.0 \cite{SM}. 
 These are not far from the value 
 $\zeta$$=$$2$$-$$1/4$$=$$1.75$ at the Kosterlitz-Thouless 
 (KT) transition \cite{KT1973}. 
 The real space correlation $\chi_{ij}(\rr)$$=$
 $\langle \phi_{i,\rr}\phi_{j,{\bm{0}}}\rangle$ 
 shows an exponential decay above $T^*$. 
 Figure \ref{fig:3}(e) shows that its correlation 
 length $\xi$ is well fitted by the exponential 
 divergence typical in the case of the KT transition 
 as $\xi\propto\exp[{\rm const.}/\sqrt{T-T^*}]$ 
 \cite{SM}. In addition, the data of $\mathcal T^+$ 
 indicate that  all the configurations of 
  $\{\eta_\ell\}$
 are realized, which leads to a relation 
 $\langle \phi'_+(\qq_1)\phi'_+(\qq_2)\phi'_+(\qq_3)\rangle$$=$
 $-\langle \phi'_+(\qq_1)\phi''_+(\qq_2)\phi''_+(\qq_3)\rangle$. 
 Here, $\phi'$($\phi''$) denotes the real 
 (imaginary) part of $\phi$. In this sense 
 the phases $\{\eta_\ell\}$ are not locked \cite{SM}.

\textit{Discussions.---}Let us discuss how to detect 
the predicted quadrupolar orders.  Below, we assume that 
the system has a weak three-dimensionality with 
a uniform stacking of $\bm{\phi}_{\rr}$ along the $z$ 
axis and the IC-TQ states have true long-range orders.  
The most direct and powerful method effective for the 
quadrupole orders is the resonant x-ray 
scattering experiments \cite{Ament2011-aj,Groot2005-cc}. 
The sTQ order is identified by observing disordered sites 
and one can use nuclear magnetic or quadrupole resonance 
for that purpose if appropriate nuclei exist as has 
been discussed for UNi$_4$B \cite{Ishitobi2023UNi4B}. 
One can also detect free quadrupole moments at the 
disordered sites by ultrasonic experiments.  
When  quadrupole order parameter spatially 
modulates \cite{Onimaru2005-db}, an electric polarization 
field ${\bm P}(\rr)$ is induced in proportion to 
$\bm{\pi}(\rr)\equiv (\partial_y\phi_{u,\rr}+\partial_x\phi_{v,\rr}, \partial_x\phi_{u,\rr}-\partial_y\phi_{v,\rr},0)$ 
and it is detectable. See a recent observation of 
polar skyrmions in  PbTiO$_3$/SrTiO$_3$ \cite{McCarter2022-tw}.

When the system has itinerant electrons,  
they feel a moir\'e of ordered quadrupoles. This should 
be reflected in the band structure and Moir\'e minibands 
are recently observed in twisted bilayer WSe$_2$ by the 
resonant inelastic light scattering experiment 
\cite{Saigal2024-fv}. Since the IC-TQ order lacks 
the local inversion symmetry, 
one expects that the spin-orbit coupling is activated, and 
it transcribes the orbital texture into a 
moir\'e of magnetizations $\bm{M}(\rr)$. It is  
induced by electric current $\bm j$ via the 
interaction energy \vspace{-0.7cm}\\
\begin{align}
\Delta E=g_1 ({\bm M}_{\qq} \times {\bm{\pi}}_{-\qq} )_zj_z
+g_2 M_{z\qq}( {\bm{\pi}}_{-\qq}\times \jj)_z,
\end{align}
\vspace{-0.7cm}\\
where $g_{1}$ and $g_2$ are constants. 
$\bm{\pi}_{\qq}$ and $\bm{M}_{\qq}$ are the Fourier 
transform of $\bm{\pi}(\rr)$ and $\bm{M}(\rr)$, respectively. 
In this Letter we have not considered magnetic degrees 
of freedom such as magnetic dipole $M_z(\rr)$ perpendicular 
to the triangular plane, and octupole 
moments $\sim M_x(\rr) [M_x^2(\rr) \MINUS 3 M_y^2(\rr)]$ or
$M_y(\rr)[M_y^2(\rr) \MINUS 3 M_x^2(\rr)]$.  
 If they exist, 
 they may order around the cores of quadrupole vortices 
 and the overall configuration is  
a hybrid quadrupole-dipole/octupole skyrmion(-like) 
lattice. 
Such a dipole-quadrupole skyrmion has been proposed 
for a spin-1 system with an elevated 
SU(3) symmetry \cite{Mikushina2002-vb}. 
See also recent analyses
\cite{PhysRevD.103.065008,PhysRevB.106.L100406,Remund2022-nd,Zhang2023-ht,Hayami2024-px}.  
Entropy release of the magnetic degrees of freedom is 
large at the sites with disordered quadrupole 
and it helps such exotic phases to persist even 
at low temperatures, as demonstrated 
for multiple-$\qq$ multipole orders
\cite{Ishitobi2021,Ishitobi2023UNi4B,Hayami2023-yx,Zhang2024-td}.
Searching for such states is one of fascinating directions in future. 

Searching materials exhibiting the orbital moir\'e is interesting. To the best of our knowledge, no such materials have been known so far. Promising local degrees of freedom are a set of partially-filled $\{2xy,x^2-y^2\}$ type $d$-electrons with the excited $3z^2-r^2$ orbital or $\{yz,zx\}$ dominated non-Kramers doublet \cite{Paramekanti2020-ae} in $d^2$ configurations in triangular systems \cite{SM}. Several non-Kramers $f$-electron systems are also interesting: an orbital triple-$\qq$ order was reported for the stacked triangular-lattice compound UPd$_3$ \cite{Mc_Ewen_a1995-cm} similar to the sTQ, while IC orbital orders were reported for the cubic compound PrPb$_3$ \cite{Onimaru2005-db}. We hope discovery of compounds exhibiting exhibiting various types of triple-$\qq$ physics through material search, paying special attention to the above-mentioned local degrees of freedom in triangular systems.

\textit{Summary.---}We have proposed a minimal quadrupole 
model which realizes orbital moir\'e texture on the 
triangular lattice. This model 
exhibits various triple-$\qq$ orders and 
they are stabilized by the cubic anisotropy that is 
forbidden in magnetic systems due to the time-reversal 
symmetry. We have shown 
the emergence of incommensurate triple-$\qq$ states 
that exhibits an orbital moir\'e texture with 
quasi-long-range order.  
We expect our results pave a basis for 
the future investigation for 
nonmagnetic multiple-$\qq$ and moir\'e physics.

\vspace{0.6cm}

\textit{Acknowledgement.}---This work was supported 
by JSPS KAKENHI (Grant Nos. JP21H01031, JP23K20824,  
and JP23H04869).

\vspace{3cm}

\appendix

\section{Eigenmodes of exchange interaction}
Let us analyze the exchange coupling $\mathcal H_{\rm ex}$ in Eq.~(3) in the main text. The Fourier bases $\bm \phi(\qq)$ are defined as  
\begin{equation}
	\bm\phi(\qq)=\frac{1}{N}\sum_\rr e^{-i\qq\cdot \rr} \bm\phi_\rr,\quad \bm\phi_\rr=\sum_\qq e^{i\qq\cdot \rr} \bm\phi(\qq), \label{FT}
\end{equation}
where $N$ is the site number.   
$\mathcal H_{\rm ex}$ is written with them as  
\begin{subequations} 
\begin{align}
	\mathcal{H}_{\rm ex}=&\frac{N}{2}\sum_{\qq } {\bm{\phi}}^{\dag}(\qq) \cdot \hat{\mathcal J}_\qq
{\bm{\phi}}(\qq),\quad {\bm{\phi}}(\qq)\equiv \begin{bmatrix}
	\phi_u(\qq)\\[2mm]
	\phi_v(\qq)
\end{bmatrix},
\end{align}
with 
\begin{align}
\hat{{\mathcal J}}_\qq & \equiv
	\begin{pmatrix}
	J\gamma_\qq+K\gamma'_\qq & K \zeta_\qq\\[2mm]
	K \zeta_\qq & J\gamma_\qq-K\gamma'_\qq \\	
\end{pmatrix},\\
\gamma_\qq &\equiv 2\cos q_x+4\cos\frac{q_x}{2}\cos\frac{\sqrt{3}q_y}{2},\\
\gamma'_\qq &\equiv 2\cos q_x-2\cos\frac{q_x}{2}\cos\frac{\sqrt{3}q_y}{2},\\
\zeta_\qq &\equiv 2\sqrt{3}\sin\frac{q_x}{2}\sin\frac{\sqrt{3}q_y}{2}, 
\end{align}
\end{subequations}
where the $\qq$-sum runs over the first Brillouin zone (1BZ) of the triangular lattice. 
 The eigenvalues of $\hat{\mathcal J}_{\qq}$ are denoted as $J_\qq^\pm$ and are  calculated as
\begin{eqnarray}
	J_\qq^{\pm} = J\gamma_\qq \pm |K|\sqrt{{\gamma'_\qq}^2+\zeta_\qq^2}.
\end{eqnarray}

Let $\qq^\star$ denotes the wavevector where the eigenvalue $J_{\qq}^-$ has the negatively largest value. For $J>0$, one easily finds that $\qq^\star$ locates at $(\Delta_{\rm M},2\pi/\sqrt{3})$ or its equivalent positions in the 1BZ, and the parameter $\Delta_{\rm M}$ is 
\begin{align}
	\Delta_{\rm M}=\begin{cases}
		0 \quad \quad &\displaystyle{\Big(|K|>\frac{2J}{5}>0\Big)}\\[2mm]
		\displaystyle{2\cos^{-1}\left[\frac{2J+|K|}{4(J-|K|)}\right]}\quad \quad &\displaystyle{\Big(|K|\le \frac{2J}{5}\Big)}
	\end{cases},
\end{align}
and $\qq^\star = (\tfrac{2\pi}{3},\frac{2\pi}{\sqrt{3}})=\qq_{{\rm K}'}$ at $|K|=0$. 
When the system undergoes a second-order phase transition to an ordered state, $\qq^\star$ is expected to be its ordering vector. Note that the $\qq^\star$ moves from $\pm \qq_{\rm K'}$ to the M point $\bm{Q}_1=(0,\frac{2\pi}{\sqrt{3}})$ or its equivalent one along the 1BZ boundary as $|K|$ increases. The eigenvector at $\qq^\star$ corresponds to the order parameter of the incommensurate states. There are six independent wave vectors: $\pm\qq_\ell^\star (\ell=1,2,3)$: 
\begin{subequations} 
\begin{eqnarray}
\qq _1^\star &=&\left(0,\tfrac{2\pi}{\sqrt{3}}\right)+\Delta_\mathrm{M}(1,0)\equiv \bm{Q}_1+\Delta_\mathrm{M} \hat{\bm e}_{0},\\
\qq _2^\star &=&\left(\pi,-\tfrac{\pi}{\sqrt{3}}\right)+\Delta_\mathrm{M} \left(-\tfrac{1}{2},-\tfrac{\sqrt{3}}{2}\right)\equiv \bm{Q}_2+\Delta_\mathrm{M} \hat{\bm e}_{2},\ \ \ \\
\qq _3^\star &=&\left(-\pi,-\tfrac{\pi}{\sqrt{3}}\right)	+\Delta_\mathrm{M} \left(-\tfrac{1}{2},\tfrac{\sqrt{3}}{2}\right)\equiv \bm{Q}_3+\Delta_\mathrm{M} \hat{\bm e}_{1},\ \ \ 
\end{eqnarray}
\end{subequations}
with $-2\pi/3 \le \Delta_{\rm M} \le 2\pi/3$. The unit vectors are defined as $\hat{\bm e}_{\ell}\equiv \hat{\bm e}(\ell \omega)$, where $\hat{\bm e}(\theta) \equiv (\cos\theta,\sin\theta)$ and $\omega\equiv \frac{2\pi}{3}$. See Eq.~(1) in the main text.  The eigenvectors $\hat{\bm{v}}_\ell^\pm$ satisfy 
\begin{align}
	\hat{\mathcal J}_{\qq_\ell^\star}\bm{v}^\pm_\ell&=\left\{J\gamma_{\qq^\star}\mp 2K \left[\cos \Delta_{\rm M} +\cos\left(\frac{\Delta_{\rm M}}{2}\right) \right]  \right\}\bm{v}^\pm_\ell,\label{eq:eigenval-eq}
\end{align}
where $\cos \Delta_{\rm M}+\cos(\Delta_{\rm M}/2)\ge 0$. 
\begin{subequations} 
They are given by 
\begin{alignat}	{2}
 \bm{v}^-_1 &=(1,0)^{\rm T}=\hat{\bm e}_0,\quad &
&\bm{v}^+_1  =(0,1)^{\rm T}=\hat{\bm e}(\FRAC{3\omega}{4}),
\\
\bm{v}^-_2 &=(-\FRAC{1}{2},\FRAC{\sqrt{3}}{2})^{\rm T} = 
\hat{\bm e}_1,\quad &
&\bm{v}^+_2=-(\FRAC{\sqrt{3}}{2},\FRAC{1}{2})^{\rm T} = 
\hat{\bm e}(\tfrac{7\omega}{4}),
\\
\bm{v}^-_3 &=-(\FRAC{1}{2},\FRAC{\sqrt{3}}{2})^{\rm T} = 
\hat{\bm e}_2,\quad & 
&\bm{v}^+_3=(\FRAC{\sqrt{3}}{2},-\FRAC{1}{2})^{\rm T} = 
\hat{\bm e}(\FRAC{11\omega}{4}).
\end{alignat}
\end{subequations}
The eigenmodes of the leading instability for $K>0$ are $\bm{v}^+_\ell$'s, while $\bm{v}^-_\ell$'s for $K<0$. Note that the sign of $2K$ in the RHS of Eq.~(\ref{eq:eigenval-eq}) is not ``$\pm$'' but represents the leading mode corresponding to sgn$(K)$.

For $J<0$, one can find
\begin{equation}
	\qq^\star=\begin{cases}
		{\bf 0} \quad \quad &(|K|<-2J)\\
		{\bm Q}_\ell \quad \quad (\ell=1,2,3)\ &(|K|>-2J)\\		
	\end{cases}.
\end{equation}
Note that $J_{\qq^\star}^+=J_{\qq^\star}^-$ for $\qq^\star=\bf 0$ and the eigenvectors are arbitrary. Typical examples of the dispersion  $J_\qq^\pm$ are shown in Fig.~\ref{fig:1}.


\begin{figure*}[t!]
\includegraphics[width=0.9\textwidth]{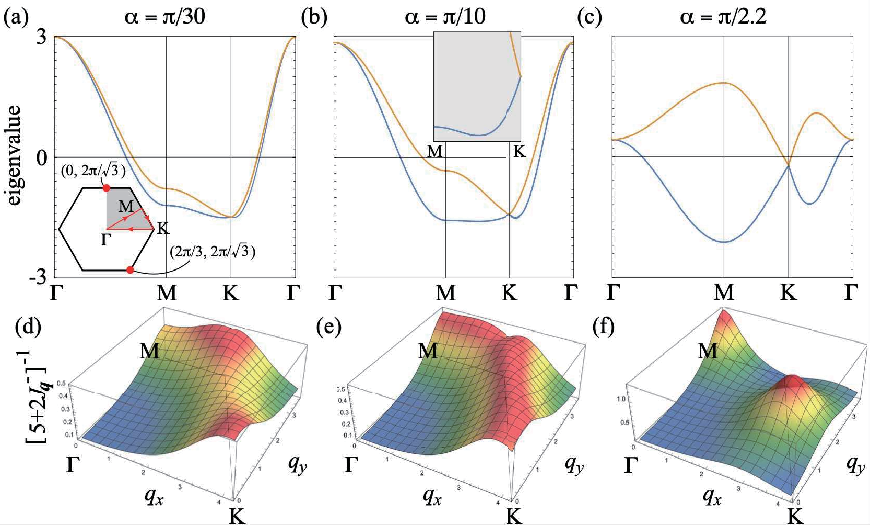}
\caption{
(a)-(c) 
Two eigenvalues $J_\qq^-$ (blue) and $J_\qq^+$ (red) 
along the high symmetry axes in the 1BZ.  
The parameters are $(J,K)=(\cos\alpha,\sin\alpha)$. 
The inset in (a) shows the 1BZ, while  
the inset in (b) is a zoom up of the region between the M and the K points. 
(d)-(f) 
$1/(5+2J^-_{\qq})$ for the data in (a)-(c), respectively.  
}
\label{fig:1}
\end{figure*}


\section{Landau Free energy}
In this section, we will derive
 the Landau free energy 
 for several phases appeared in the main text.  
 We will not show the details of  minimization procedure 
 for simplicity but the line of discussion is similar to 
 that for cubic systems \cite{Tsunetsugu2021,Hattori2023-pj}. 

The free energy up to the fourth order in $\bm{\phi}$'s is given as 
\begin{align}
	F&=\frac{N}{2}\sum_\qq \bm{\phi}^\dag(\qq) \left(a_0+\hat{\mathcal J}_\qq\right)\bm{\phi}(\qq)
	\nonumber\\
	&+\sum_\rr\Big[\lambda\Big(3\phi_{u,\rr}^2-\phi_{v,\rr}^2\Big)\phi_{v,\rr} +c|\bm{\phi}_{\rr}|^4+ \cdots\Big], \label{eq:Forig}
\end{align}
where $a_0=\bar{a}_0T$ with $T$ being temperature, and $\bar{a}_0,\ \lambda,\ c$ are positive constants. 

In the following, we assume the ordering vectors are $\qq^\star_{\ell}  (\ell=1,2,3)$, which are either  commensurate $\bm{Q}_\ell$ or incommensurate (IC) ones on the 1BZ boundary. Taking into account the mode-mode coupling terms arising from the cubic term, one finds that 
$\bm\phi(\bf 0)$  couples with $\bm{\phi}(\qq^\star_\ell)$ in the linear order in $\bm{\phi}(\bf 0)$. The other modes in the cubic term just appear at least in the quadratic orders and are safely ignored. As for the fourth-order mode-mode coupling terms, they are subleading compared to the cubic term proportional to $\lambda$. Indeed, when $|\bm{\phi}(\qq_\ell^\star)|$ is $\ell$-independent, {the contributions proportional to $c$} vanish.

 First, let us consider the commensurate cases with $\qq^\star=\bm Q_\ell$: $\bm{\phi}(\bm{Q}_\ell)=\sigma_\ell\phi_\ell {\bm v}^s_\ell$ ($\sigma_\ell=\pm 1$, $\phi_\ell\ge 0$, $s=+$ or $-$), which couples with $\bm{\phi}(\bf 0)$. The free energy $F=F_s$ is given by
\vspace{-6pt}
\begin{align}
\frac{F_s}{N}
&=
\frac{1}{2}\sum_{\ell=1,2,3} (a_0+J^-_{\bm{Q}_\ell}){\phi_\ell}^2
-6\lambda\sigma_1\sigma_2\sigma_3\phi_1\phi_2\phi_3\delta_{s,+} 
\nonumber\\
&+
\frac{c}{2}
\Bigl[ 
3(\phi_1^2+\phi_2^2+\phi_3^2)^2 
- \!\!\! \sum_{\ell=1,2,3} \!\! \phi_\ell^4 
\Bigr]
-\frac{3s\lambda}{2}\bm{D}\cdot\bm{\phi}(\bm{0}). \label{eq:Fs}
\end{align}
Here, we have kept terms linear in $\bm{\phi}({\bf 0})$ and 
\begin{align}
\bm{D}\equiv \begin{bmatrix}
	\sqrt{3}(-\phi_2^2+\phi_3^2)\\[1mm]
	2\phi_1^2-\phi_2^2-\phi_3^2	
\end{bmatrix}=2\sum_{\ell=1,2,3}\phi_\ell^2\bm{v}^+_{\ell}.	\label{eq:Dvector}
\end{align}
{This is represented with $\bm{v}^+_\ell$ irrespective of the sign of $s$. }

When the leading ordering vectors are IC $\qq^\star_\ell$'s, 
they are distinct from $-\qq^\star_\ell$ and we replace $\bm{\phi}(\bm{Q}_\ell)$ by  
\begin{align}
	\bm{\phi}(\qq_\ell^\star)=\phi_\ell e^{i\eta_\ell}\bm{v}_\ell^s,\quad 
	\bm{\phi}(-\qq_\ell^\star)=[\bm{\phi}(\qq_\ell^\star)]^*. \label{eq:DefIC}
\end{align}
Then, the free energy $F=F_s^{\rm IC}$ becomes
\begin{align}
	\frac{F_s^{\rm IC}}{N}&=\sum_{\ell=1,2,3} (a_0+J^{-}_{\qq^\star_\ell}){\phi_\ell}^2
	-12\lambda\phi_1\phi_2\phi_3\cos\eta_{\rm tot}\delta_{s,+}\nonumber\\
	&+6c(\phi_1^2+\phi_2^2+\phi_3^2)^2-3s\lambda\bm{D}\cdot\bm{\phi}({\bf 0}),
\end{align}
where $\eta_{\rm tot}=\eta_1+\eta_2+\eta_3$.

\subsection{Single-$q$ states}
We now consider the free energy density of the single-$\qq$ states. 
 The results for commensurate and IC cases  are
\begin{align} 
	\frac{F_{s\ell}}{N}&=\frac{1}{2}(a_0+J^{-}_{{\bm Q}_\ell})\phi_\ell^2+c\phi_\ell^4-3s\lambda\phi_\ell^2\bm{v}^+_{\ell}\cdot \bm{\phi}(\bf 0),\label{Fsl-M}\\
	\frac{F^{\rm IC}_{s\ell}}{N}&=(a_0+J^{-}_{{\qq}^\star_\ell})\phi_\ell^2+6c\phi_\ell^4-6s\lambda\phi_\ell^2\bm{v}^+_{\ell}\cdot \bm{\phi}(\bf 0),\label{Fsl-IC}
\end{align}
respectively. Here, we have used Eq.~(\ref{eq:DefIC}) 
for deriving Eq.~(\ref{Fsl-IC}). 
We rescale the order parameters as 
$\phi_\ell=\varphi_\ell/\sqrt{2}$ in $F^{\rm IC}_{s\ell}$. 
Then, we obtain
\begin{align}
	\frac{F^{\rm IC}_{s\ell}}{N}&=\frac{1}{2}(a_0+J^{-}_{{\qq}^\star_\ell})\varphi_\ell^2+\frac{3c}{2}\varphi_\ell^4-3s\lambda\varphi_\ell^2\bm{v}^+_{\ell}\cdot \bm{\phi}({\bf 0}).\label{Fsl-IC2}
\end{align}
Note that the fourth-order term in 
$F_{s\ell}^{\rm IC}$ 
[Eq.~(\ref{Fsl-IC2})] is larger than that in $F_{s\ell}$ 
[Eq.~(\ref{Fsl-M})], and this represents the commensurate locking. See Appendix \ref{sec:CL}. 
It is clear that once $\varphi_\ell$ or $\phi_\ell$ is finite, 
the ferroic moment $\bm{\phi}({\bf 0})$ is induced 
along the direction of $s\bm{v}^+_{\ell}$ 
for $\lambda>0$, which we have named canted SQ state (cSQ).  
This kind of induction of uniform moment is common 
in the quadrupole systems of ``E''-type 
(two-dimensional irrep.)\cite{Hattori2014,Hattori2016,Tsunetsugu2021,Hattori2023-pj}.

\subsection{Symmetric triple-$q$ state for $K>0$}\label{sec:STQ}
Let us discuss the symmetric triple-$\qq$ (sTQ) state 
with $\bm{Q}_\ell$'s for $\ K>0$. 
As explained in the main text, this state is a 
symmetric superposition of the three modes 
with the equal magnitude $\phi_\ell=\phi/2$

\begin{eqnarray}
	\bm{\phi}_\rr&=&\frac{\phi}{2}
	\sum_{\ell=1,2,3} \sigma_\ell \cos({\bm Q}_\ell \cdot \rr) {\bm v}_\ell^+.
\end{eqnarray} 
Note that $\sigma_1\sigma_2\sigma_3$ in Eq.~(\ref{eq:Fs}) takes
 the value 1 in order to maximize the energy gain. 
The local order parameters $\bm{\phi}_\rr$ are eigher  
${\bf 0}$ or one of $\phi\bm{v}^+_{\ell} (\ell=1,2,3)$ and this is a four-sublattice order with one disordered sublattice. For $\sigma_1=\sigma_2=\sigma_3=1$, the four-sublattice configuration is that with 
$
\bm{\phi}_{(0,0)}={\bf 0}$, 
$\bm{\phi}_{(1,0)}=\hat{\bm e}(\frac{\pi}{2})$, 
$\bm{\phi}_{(1/2,\sqrt{3}/2)}=\hat{\bm e}(\frac{7\pi}{6})$, and 
$\bm{\phi}_{(3/2,\sqrt{3}/2)}=\hat{\bm e}(\frac{11\pi}{6})
$. See Fig.~1(c) in the main text. There are three other domains and they are realized 
by shifting the position $\rr \to \rr -\tfrac{1}{2}(\sigma_1+2\sigma_2+\sigma_3)\hat{\bm e}(0)-\tfrac{1}{2}(2\sigma_1+\sigma_2+\sigma_3)\hat{\bm e}(\omega/2)$.

The total free energy density for the sTQ state is 
\begin{align}
		\!\!\!\!\!\!\frac{F_{\rm sTQ}}{N}=\frac{1}{2}\sum_{\ell=1,2,3}(a_0+J_{\bm{Q}_\ell}^-)\left(\frac{\phi}{2}\right)^2
	-\frac{3\lambda}{4}\phi^3 +\frac{3c}{4}\phi^4.	\label{FsTQ0}
\end{align}
Here, we have used the relation $\bm{\phi}(\bm{Q}_\ell)=\sigma_\ell\phi \bm{v}_{\ell}^+/2$, $\bm{D}=\bm{0}$ when $\phi_\ell=\phi/2$ and 
\begin{align}
	J_{\bm{Q}_\ell}^-=-2J-4K, \quad (K>0).
\end{align}
The fact that one sublattice is disordered is apparent in the factor $3/4$ of the fourth order term in Eq.~(\ref{FsTQ0}). 
In order to discuss the stability against 
the single-$\qq$ order, it is useful to rescale $\phi$ 
such that the quadratic term has the same form  
as in Eq.~(\ref{Fsl-M}): $\phi=\frac{2}{\sqrt{3}}\varphi$. 
The result is 
\begin{align}
		\frac{F_{\rm sTQ}}{N}=\frac{1}{2}(a_0+J_{\bm{Q}}^-)\varphi^2
	-\frac{2\lambda}{\sqrt{3}}\varphi^3 +\frac{4c}{3}\varphi^4.	
	\label{eq:FsTQ}
\end{align}
The presence of $\varphi^3$-term demonstrates that the transition into the sTQ state 
is first order and its transition temperature is 
higher than that of the single-$\qq$ state 
in the simple Landau analysis. This situation is similar to the cases 
in cubic systems \cite{Tsunetsugu2021,Hattori2023-pj}.
However, as demonstrated in the main text, this is not the case when the fluctuation effects are taken account, and the transition becomes second order.

\subsection{Asymmetric triple-$q$ state for $K>0$}
The sTQ state is expected to be unstable at 
lower temperatures due to the presence of 
disordered sites. 
Here, we will sketch 
the symmetry breaking from the sTQ to the asTQ 
states. Parameterizing $\bm{\phi}({\bf 0})$$=$
$\phi_0\hat{\bm{e}}(\theta_0)$ and 
$\bm{\phi}({\bm Q}_\ell)=\Phi_\ell\hat{\bm{e}}(\theta_\ell)$, 
one finds that the cubic anisotropy energy in Eq.~(\ref{eq:Forig}) contains the following term  
\begin{align}
	3\lambda \sum_{\ell=1}^3\phi_0\Phi_\ell^2 \sin(\theta_0+2\theta_\ell).
\end{align}
Let us consider small asymmetries in the three modes 
$\Phi_\ell=\Phi+\delta_\ell$ and 
$\theta_\ell=(1-\tfrac{\ell}{4})\omega+\epsilon_\ell$.  
Then, this term becomes 
\begin{align}
	&3\lambda\phi_0\Phi\Big\{
	  \sin\theta_0  \left[-2 \delta _1+\delta _2+\delta _3-\sqrt{3} \left(\epsilon_2-\epsilon_3\right) \Phi \right]\nonumber\\
	&+ \cos\theta_0  \left[\sqrt{3} (\delta _2-\delta _3)+\left(-2 \epsilon_1+\epsilon_2+\epsilon_3\right) \Phi \right]
	\Big\},
\end{align} 
up to the linear order in $\delta_\ell$ or
 $\epsilon_\ell$. This shows that the asymmetries  
in the three modes induce $\phi({\bf 0})$. 
Note that the $\phi({\bf 0})$ is coupled with not only 
 the magnitude asymmetry $\delta_\ell$
 but also the angle asymmetry $\epsilon_\ell$. 
 This means that in general both the amplitude and the angle 
 of $\bm{\phi}(\bm{Q}_\ell)$ 
 in the asTQ state deviates from those in the sTQ state. 
 Similar discussion has been done in Ref.~\onlinecite{Hattori2023-pj}.

\subsection{Incommensurate triple-$q$ state}
Let us consider IC triple-$\qq$ states. 
For simplicity, we restrict ourselves to the 
symmetric state for $K>0$.  
Similarly to the case of the sTQ state, 
$\bm{\phi}_{\rr}$ is expressed as  
\begin{eqnarray}
	{\bm \phi}_{\rr} &=& \sum_{\ell=1,2,3} \frac{2\varphi}{\sqrt{6}} \bm{v}^+_\ell \cos(\qq^\star_\ell\cdot \rr+\eta_\ell).
\end{eqnarray}
This corresponds to $\bm{\phi}(\qq_\ell^\star)=\sqrt{N/6}~\varphi\ e^{i\eta_\ell}\bm{v}^+_{\ell}$ and the free 
energy density reads as 
\begin{align}
	\!\!\!\!\!\!\frac{F_{\rm IC-sTQ}}{N}=\frac{1}{2}(a_0+J_{\qq^\star}^-)\varphi^2
	-\frac{2\lambda}{\sqrt{6}}\varphi^3\cos\eta_{\rm tot} 
	+\frac{3c}{2}\varphi^4.	
	\label{eq:FICtQ}
\end{align}
Here, in order to maximize the energy gain, $\eta_{\rm tot}=2\pi  \times $(integer) should be satisfied. Comparing this to the result of the commensurate 
case, one observes the commensurate locking, i.e., 
the third- and forth-order terms have larger or negatively smaller coefficients than those in Eq.~(\ref{eq:FsTQ}). {See also Appendix \ref{sec:STQ}}. 

\subsection{120$^\circ$ antiferroic state}
For $K=0$, the ordering vectors locate 
at the K or K$'$ points: 
$\qq_{\rm K_1}=2\omega \hat{\bm e}_0$, 
$\qq_{\rm K_2}=2\omega\hat{\bm e}_1$, 
$\qq_{\rm K_3}=2\omega\hat{\bm e}_2$, where the 
eigenvalues $J_{\qq_{{\rm K}_{\ell}}}^{\pm}$ are 
degenerate. Thus, the directions of 
$\bm{\phi}({\qq_{{\rm K}_{\ell}}})$ are not determined 
by the quadratic part of the free energy. 
The ordering configuration 
is chosen by minimizing the local third-order term which favors 
the three directions: $\hat{\bm e}(3\omega/4)$, 
$\hat{\bm e}(7\omega/4)$, and $\hat{\bm e}(11\omega/4)$. 
Such a 120$^\circ$ AF state is realized as a special case of the 
continuously degenerate 120$^\circ$ structure. 
One realization is 
\begin{eqnarray}
	\bm{\phi}_\rr 
	&=&\frac{\phi}{\sqrt{2}}\left[ 
	\frac{e^{i\qq_{\rm K_1}\cdot \rr}}{\sqrt{2}}\begin{pmatrix}
		i\\
		1
	\end{pmatrix}
	+{\rm c.c}\right]=\phi\begin{bmatrix}
		-\sin(\qq_{\rm K_1}\!\!\cdot\!\rr)\\
		\cos(\qq_{\rm K_1}\!\!\cdot \!\rr)
	\end{bmatrix}.\ \ \ \ \ \ 
\end{eqnarray} 
Since $\phi({\qq_{\rm K_1}})=\phi/\sqrt{2}$, the free energy is 
\begin{eqnarray}
	\frac{F_{120^\circ}}{N}= \frac{1}{2}\left(a_0-3J\right) \phi^2-\lambda\phi^3+c\phi^4. \label{eq:F120}
\end{eqnarray}
The presence of the $\phi^3$-term implies 
that the transition to a 120$^\circ$ 
AF state is first order. 
{Since the 120$^\circ$ 
AF states realize six 
domains, it is expected that the phase transition is described by the 6-state Potts class.
Thus, the transition is expected to be first order in two dimensions\cite{Wu1982-sm}. Indeed, the Monte Carlo data support this simple Landau analysis.}

\subsection{Ferroic state}
The ferroic quadrupolar state is simply given by 
$\bm{\phi}_{\bm r}=\varphi\ \bm{v}^+_{\ell} (\ell\in \{1,2,3\})$. The direction of $\bm{\phi}_{\bm r}$ is determined by the cubic term $\lambda(>0)$ and is three-fold degenerate. The free energy density is 
\begin{align}
	\frac{F_{\rm F}}{N}=\frac{1}{2}(a_0+6J)\varphi^2-\lambda \varphi^3 +c\varphi^4.\label{eq:FFQ}
\end{align}
Again, this predicts a first-order transition. 
However, the results of our numerical simulation in the main text reveal it is second order, and this is consistent with the universality class of the 3-state Potts model in two dimensions\cite{Wu1982-sm}.

{
\subsection{Microscopic estimation of $\lambda$}
To help readers' understanding about the cubic coupling $\lambda$ in the Landau free energy (\ref{eq:Forig}), we sketch its derivation on the basis of a few  representative microscopic models. 

Consider a $d$-electron system with partially filled $d_{xy}$ and $d_{x^2-y^2}$ orbitals, and its $d^1$ configuration, ignoring the spin part. In addition to these orbitals, we include  an excited $d_{3z^2-r^2}$ orbital with the energy $\Delta$ above that for the degenerate $\{d_{xy},d_{x^2-y^2}\}$ doublet. Then, the E$_2$ orbital operators $\{u,v\}\sim \{2xy,x^2-y^2\}$ are represented as
\begin{align}
	u=\begin{bmatrix}
	0 & a & b\\[1mm]
	a & 0 & 0\\[1mm]
	b & 0 & 0\\[1mm]
\end{bmatrix},\quad 
v=\begin{bmatrix}
	a & 0 & 0\\[1mm]
	0 & -a & b\\[1mm]
	0 & b & 0\\[1mm]
\end{bmatrix},\label{eq:u_v}
\end{align}
where the basis set $\{d_{xy},d_{x^2-y^2},d_{3z^2-r^2}\}$ with  $a^2+b^2=1$. Here, the orbital operators consist of two parts: the quadrupole and the hexadecapole moments. The quadrupole part is given by $\{L_{x}L_y+L_yL_x,\ L_x^2-L_y^2\}=\{\tfrac{1}{2i}(L_+^2-L_-^2),\tfrac{1}{2}(L_+^2+L_-^2)\}$, where $\{L_{\mu}\}_{\mu=x,y,z}$ are the orbital angular momentum operators for the $d$ electrons and $L_\pm=L_x\pm iL_y$. This is the part proportional to $b$. The hexadecapole part is proportional to $a$ and this is given by $\{L_{x}^3L_y-L_xL_y^3 +$ perm.$,\ (L_x^2L_y^2 + $ perm.$)-L_{x}^4-L_y^4\}=\{\tfrac{1}{2i}(L_+^4-L_-^4),-\tfrac{1}{2}(L_+^4+L_-^4)\}$. Here, ``perm." represents the sum for all possible permutations of $L_{x,y}$. The actual values of $a$ and $b$ depend on the exchange interactions. The detail of the exchange interactions fixes the ratio of the quadrupole and hexadecapole component relevant to the phase transition considered. These forms of orbital operators are essentially the same as those discussed for the cubic $\Gamma_3$ system \cite{Hattori2014}. 

Following the derivation in Ref.~\onlinecite{Hattori2014}, we can calculate the local free energy under the conjugate fields $\tilde{h}_{u}$ and $\tilde{h}_{v}$ to $u$ and $v$, respectively. Then, carrying out the Legendre transformation, we find the single-site Landau free energy as 
\begin{align}
	F_{\rm loc}\simeq \frac{\alpha(T)}{2}(\phi_u^2+\phi_v^2)+\lambda(T) (3\phi_v\phi_u^2-\phi_v^3)+\cdots.
\end{align}
Here, we have ignored trivial constant parts and $\{\phi_u,\phi_v\}$ is the vector classical field with E$_2$ symmetry  corresponding to $\{\langle u\rangle, \langle v\rangle\}$. The coefficients $\alpha(T)$ and $\lambda(T)$ are
\begin{align}
	\alpha(T)&=\frac{\Delta}{2b^2} \left[
	\frac{2+e^{-\Delta/T}}{1+(a/b)^2(\Delta/T) -e^{-\Delta/T}}\right]
	\nn\\
	&\sim \frac{T}{a^2} \quad (\Delta\gg T)
,\\
	\lambda(T)&=\frac{a\Delta}{8b^4} (2+e^{-\Delta/T})^2
	\frac{(\Delta/T)-1+e^{-\Delta/T}}{[1+(a/b)^2(\Delta/T) -e^{-\Delta/T}]^3}\nn\\
	&=\alpha^3(T) \frac{b^2}{\Delta^2}\frac{(\Delta/T)-1+e^{-\Delta/T}}{2+e^{-\Delta/T}}\nn\\
	&\sim \frac{b^2\alpha(T)}{2}
\left[\frac{\alpha(T)}{\Delta}\right]
\left[\frac{\alpha(T)}{T}\right]\propto T^2
	\quad (\Delta\gg T).
\end{align} 
Here, $T$ is temperature. 

Let us estimate the order of magnitude of $\lambda$ at the second-order transition temperature.  
First, when $\Delta\sim T_c\sim \alpha(T_c)$, one finds $\lambda(T_c)\sim T_c$. 
Thus, one needs materials with smaller $\Delta$ or larger exchange couplings in the orbital sector. Next, let us consider cases for $\Delta\gg T$. 
When $T$ is close to $T_c\sim a^2\alpha(T_c)$,
$\lambda(T_c)\sim (b/a^3)^2(T_c/\Delta)T_c/2$. 
Thus, setting $T_c\sim \Delta/10$ as a typical value for $d$-electron systems, $\lambda=(b/a^3)^2T_c/20$.
 Parameterizing as $a=\cos\psi$ and $b=\sin\psi$, one obtains $\lambda=\frac{1}{20}[\tan\psi(1+\tan^2\psi)]^2T_c$. 
This leads to $\lambda(T_c)=T_c$ for $\psi\simeq 0.3\pi$
and $0.5T_c$ for $\psi\simeq 0.28\pi$. 
It is noted that  $\lambda(T_c)\sim T_c$ when $b$ is slightly larger than $a$. We expect such a situation can be widely realized when the orbital operators have  matrix elements between the ground doublet and the excited state. For example, $\psi\sim 0.4\pi$ for the quadrupole operators in cubic $\Gamma_3$ systems\cite{Hattori2014}.

It is also interesting to investigate a non-Kramers system with $d^2$ configurations , and it has been discussed for Os based double perovskites\cite{Paramekanti2020-ae}. When the spin-orbit coupling is large such as in $5d$ systems,  crystalline-field ground state can be a doublet E$_2$ with the total angular momentum $J=2$ which has no dipole moment as in $f^2$ systems\cite{Hattori2014}.
	Such E$_2$ ground states can be obtained by diagonalizing the full ionic  Hamiltonian containing the electron-electron Coulomb repulsions, the crystalline fields, and the spin-orbit coupling. To make the discussion simple, we assume a restricted Hilbert space and antiferromagnetic Hund's couplings to realize the E$_2$ doublet ground states in the $d^2$ configurations. This can be realized by taking into account renormalization of the interaction due to the high-energy excited states\cite{Hattori2017-rg} or electron-phonon couplings\cite{Nomura2015-gi,Hoshino2017-be}. 

One may naively uses the single-particle bases $\{d_{xy},d_{x^2-y^2},d_{3z^2-r^2}\}$ as for the $d^1$ configuration above and construct $d^2$ states with the E$_2$ symmetry and the total spin $S=0$. The two-electron $E_2$ operators $\{u,v\}$ have  the same form as Eq.~(\ref{eq:u_v}), and the similar results follow. A simpler  model with finite $\lambda$ is constructed by using the single-particle bases $\{d_{yz},d_{zx}\}$, where
\begin{align}
	d_{yz\sigma}^\dag=-\frac{d^\dag_{1\sigma}+d_{-1\sigma}^\dag}{\sqrt{2}i} ,\quad 
	d_{zx\sigma}^\dag=-\frac{d^\dag_{1\sigma}-d_{-1\sigma}^\dag}{\sqrt{2}}.
\end{align}
Here, $d_{m\sigma}^\dag$ is the creation operator with the $z$ component of the orbital angular momentum projection $\ell_z=m$ and the spin $\sigma=\uparrow,\downarrow$. The $d^2$ states with E$_2$ symmetry $\{|{\rm E}_{2u}\rangle,|{\rm E}_{2v}\rangle\}\sim\{2xy,x^2-y^2\}$ are
\begin{align}
	|\rm{E}_{2u}\rangle &= \frac{d^\dag_{yz\uparrow}d^\dag_{zx\downarrow}-d_{yz\downarrow}^\dag d_{zx\uparrow}^\dag}{\sqrt{2}}|0\rangle
	=\frac{d^\dag_{1\uparrow}d^\dag_{1\downarrow}-d_{-1\uparrow}^\dag d_{-1\downarrow}^\dag}{\sqrt{2}i}|0\rangle
	,\\
	|\rm{E}_{2v}\rangle &= \frac{d^\dag_{zx\uparrow}d^\dag_{zx\downarrow}-d_{yz\uparrow}^\dag d_{yz\downarrow}^\dag}{\sqrt{2}}|0\rangle
		=\frac{d^\dag_{1\uparrow}d^\dag_{1\downarrow}+d_{-1\uparrow}^\dag d_{-1\downarrow}^\dag}{\sqrt{2}}|0\rangle,
\end{align}
and that with A$_1$ symmetry is 
\begin{align}
	|{\rm A}_1\rangle &= -\frac{d^\dag_{zx\uparrow}d^\dag_{zx\downarrow}+d_{yz\uparrow}^\dag d_{yz\downarrow}^\dag}{\sqrt{2}}|0\rangle
		=\frac{d^\dag_{1\uparrow}d^\dag_{-1\downarrow}-d_{1\downarrow}^\dag d_{-1\uparrow}^\dag}{\sqrt{2}}|0\rangle.
\end{align}
Here, $|0\rangle$ is the vacuum. It is clear that these states consist of the states with the $z$ component of the total angular momentum $J_z=\pm 2,0$. Using the bases $\{|{\rm E}_{2u}\rangle, |{\rm E}_{2v}\rangle,|{\rm A}_1\rangle\}$, we can construct E$_2$ operators $\{u,v\}$, which have the same matrix elements as Eq.~(\ref{eq:u_v}). We note that the above discussion is modified when other orbitals are hybridized.}


\section{Monte Carlo simulation}
In this section, we will present several details of Monte Carlo (MC) data not shown in the main text.  The exchange constants are parameterized as before as $(J,K)=\bar{J}(\cos\alpha,\sin\alpha)$ with $\bar{J}=1$. 


\begin{figure}[t!]
\includegraphics[width=0.45\textwidth]{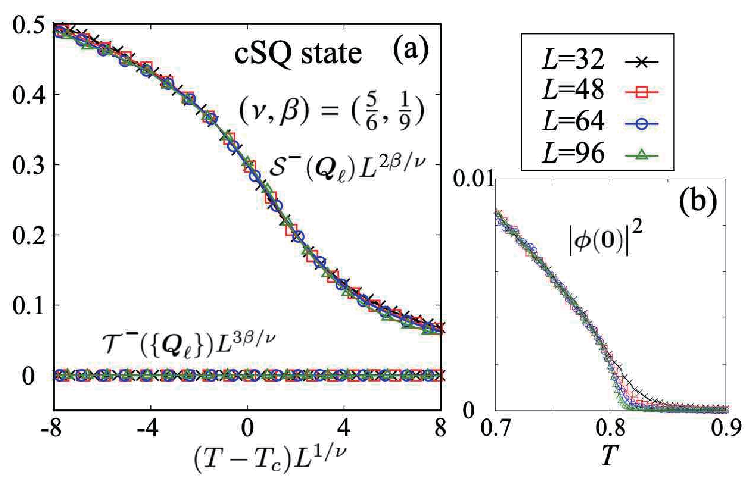}
\caption{
(a) Finite-size scaling analysis of the order parameter structure factors 
for the transition to the cSQ state. 
The used exponents are those of the P3 universality class.  
The model parameters are $\alpha=-3\pi/8$ and $\lambda=1$, 
and the transition temperature is assumed $T_c=0.8032$. 
Note that the eigenmode is $\bm v^-_\ell$ and 
the triple-$\qq$ structure factor 
$\mathcal{T}^-$ vanishes. 
(b) The squared uniform moment $|{\bm{\phi}}(\bm{0})|^2$ vs $T$.
}
\label{fig:3stPottsKm}
\end{figure}


\subsection{Three-state Potts universality}

Let us start to discuss the universality class 
of the transition to the cSQ state. 
In contrast to the Ashkin-Teller class or the 
four-state Potts (P4) of the sTQ order, it belongs 
to the three-state Potts (P3) class \cite{Wu1982-sm}. 
The transition to the F ordered state is also 
governed by the same P3 class, which we will not 
discuss here since this is a trivial one. 
As for the AF120$^\circ$ order, the transition is 
first order as expected in the six state Potts model 
in two dimensions \cite{Wu1982-sm}, 
and we will not discuss it furthermore.

Figure \ref{fig:3stPottsKm}(a) shows the finite-size 
scaling analysis of the single-$\qq$ structure 
factor $\mathcal S^-({\bm Q}_\ell)$ defined in 
the main text. The parameters are set to $\alpha=-3\pi/8$ and $\lambda=1$. For checking whether 
the triple-$\qq$ order is present, we have also calculated  
$\mathcal T^-(\{{\bm Q}_\ell \})$, but it turns out to 
be zero. This confirms the absence of triple-$\qq$ 
orders for $K<0$ and $\lambda=1$. As clearly seen 
in Fig.~\ref{fig:3stPottsKm}(a), the data are well 
scaled with the exponents of the P3 class \cite{Wu1982-sm}. The cubic 
term $\propto\lambda$ induces finite ferroic 
moments $\bm{\phi}({\bf 0})$. Figure \ref{fig:3stPottsKm}(b) 
shows the $T$ dependence of the secondary order parameter 
$|{\bm \phi}({\bf 0})|^2$. The size dependence is rather 
small below $T_c=0.8032$ and the data for 
different $L$'s collapse onto a single curve. 
This indicates $|{\bm \phi}({\bf 0})|>0$ in the ordered 
state and this is what we have called the canted 
single-$\qq$ order: cSQ. The exponents for the secondary 
order parameter is also one of interesting recent 
topics \cite{PhysRevB.109.L140408} but here we 
do not discuss the detail.

\begin{figure}[t!]
\includegraphics[width=0.45\textwidth]{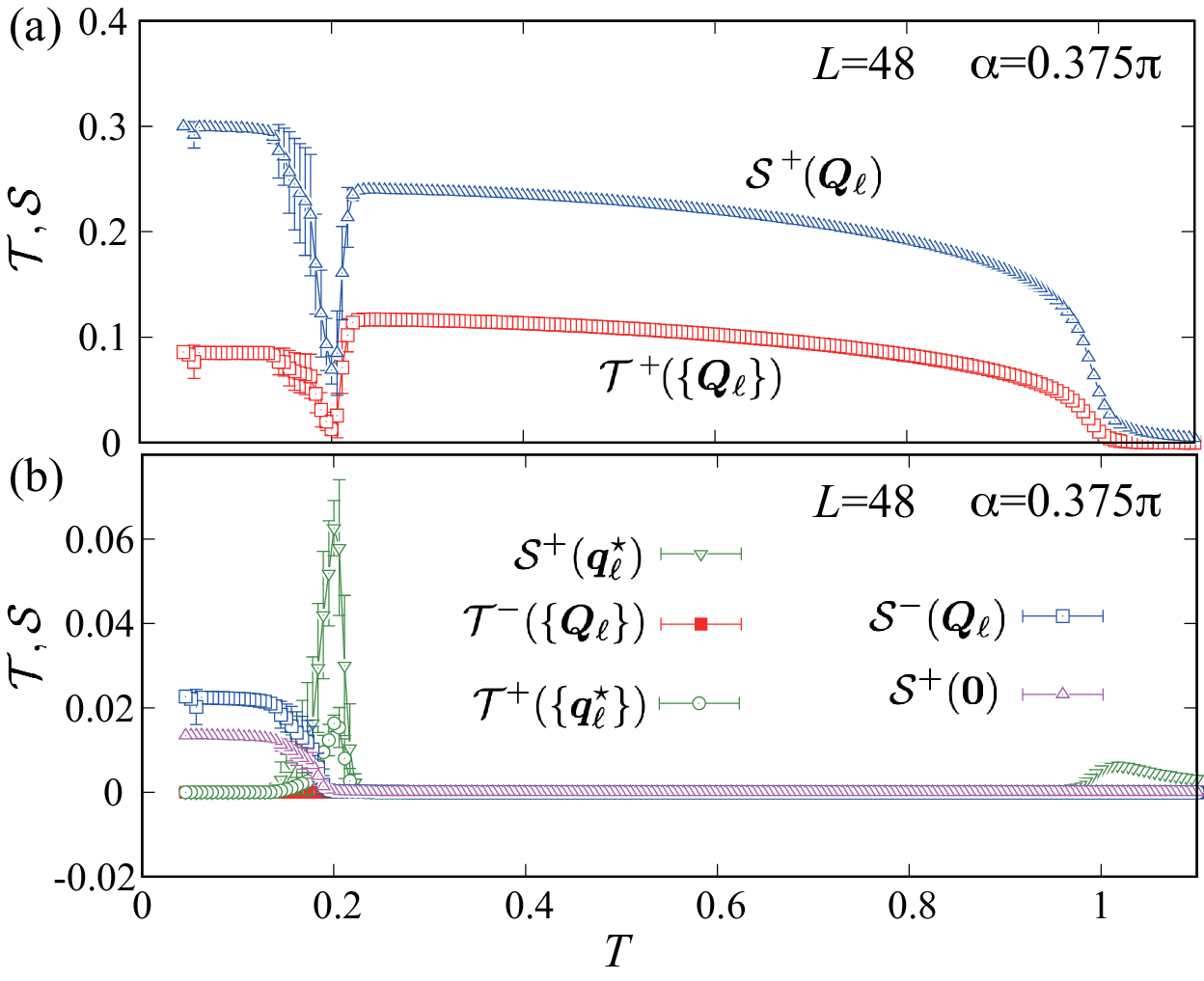}
\caption{
Temperature dependence of 
various structure factors in the system with $L=48$, $\alpha=0.375\pi$, and $\lambda=1$. (a) ${\mathcal S}^+({\bm Q}_\ell)$ and ${\mathcal T}^+(\{{\bm Q}_\ell\})$ as a function of $T$. (b) ${\mathcal S}^+({\qq}^\star_\ell)$, ${\mathcal T}^-(\{{\bm Q}_\ell\})$, ${\mathcal T}^+(\{{\qq}^\star_\ell\})$, ${\mathcal S}^-({\bm Q}_\ell)$, and ${\mathcal S}^+({\bm 0})={\mathcal S}^-({\bm 0})$ as a function of $T$. Here, ${\qq}^\star_\ell={\bm Q}_\ell\pm \Delta_{\rm M} \hat{\bm e}_{4-\ell}$ with $\Delta_{\rm M}=\tfrac{4\pi}{3}\tfrac{1}{48}$. Data are averaged over four ensembles that start from different initial configurations and are annealed for 10$^7$ MCS. Error bars show the standard deviation of these four sets.}
\label{fig:stQ-asTQ}
\end{figure}


\subsection{Phase transition from sTQ to asTQ}
We now discuss the symmetry breaking associated 
with the transition from the sTQ state to the asTQ state 
upon lowering $T$. This corresponds to the 
emergence of finite moments at the disordered 
sites in the sTQ state. Figures \ref{fig:stQ-asTQ}(a) 
and (b) show the $T$-dependence of 
 the MC data of ${\mathcal T}^+(\{ \bm{Q}_\ell \})$, 
${\mathcal S}^+(\bm{Q}_\ell)$, and ${\mathcal S}^+(\bm{0})$ 
for $\alpha=\frac{3\pi}{8}$, 
which is the same parameter set 
used in Fig.~3(b) in the main text. 
For $0.2\lesssim T\lesssim 1$, the sTQ state is 
realized, where both $\mathcal T^+(\{{\bm Q}_\ell\})$ 
and $\mathcal S^+({\bm Q}_\ell)$ are finite, 
while ${\mathcal S}^+({\bf 0})=0$. As $T$ decreases, the sTQ state is replaced by an 
IC triple-$\qq$ state around 
$T\simeq 0.2$. However, this IC state is unstable 
against the anisotropic triple-$\qq$ state (asTQ) at the 
lower temperatures. This is seen 
in the fact that $\mathcal T^+(\{{\bm Q}_\ell\})$ 
is smaller and $\mathcal S^+({\bm Q}_\ell)$ is larger 
compared with those in the sTQ state. 
Note that the asTQ state has a 
nonvanishing ${\mathcal S}^{\pm}({\bf 0})$. 
This is understood by noting that 
$\bm{D}$ in Eq.~(\ref{eq:Dvector}) is nonvanishing in the asTQ state. 
In addition to this, the 
structure factor for the non-dominant eigenmode 
${\mathcal S}^-({\bm Q}_\ell)$ is also finite, and this  indicates that the moments $\bm{\phi}(\rr)$'s rotate from the configuration in the sTQ state. See 
Fig.~\ref{fig:stQ-asTQ}(b). 
The intermediate IC-TQ state 
appears only in a narrow region, and this has not been  detected in the smaller
systems with $L\le 32$. Unfortunately, we have not been able to simulate large enough systems  to check the stability of the 
asTQ and IC-TQ states in this low temperature range. The ordering vectors $\qq^\star$'s are  near 
the M points and the triple-$\qq$ configuration and 
its moir\'e are slightly different from those discussed 
in the main text. We will briefly discuss this point 
at the end of Appendix \ref{sec:IC-tQ}.

\subsection{IC-triple-$q$ state}\label{sec:IC-tQ}
In this section, we study the IC triple-$\qq$ states, focusing on the typical parameter set 
 used in the main text: $(J,K)=(\cos\alpha,\sin\alpha)$ with $\alpha=0.0938\pi$, $\lambda=1$, and $L=50$. See Fig.~3(c) in the main text. 


\begin{figure}[t]
\includegraphics[width=0.45\textwidth]{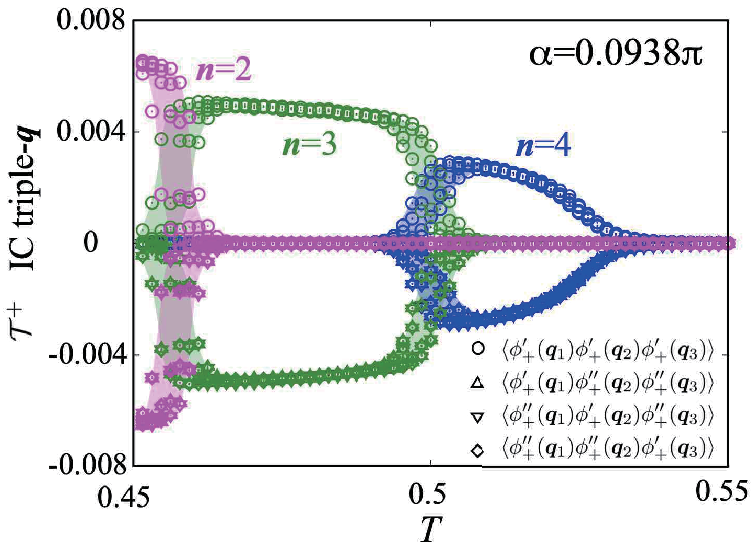}
\caption{Temperature dependence of the structure factor $\mathcal T^+$ for $\alpha=0.0938\pi$, $L=50$, and $\lambda=1$. $n$ represents the distance from the K$^\prime$ point: $\qq_\ell^\star=\qq_{{\rm K}^\prime}-\Delta_n \hat{{\bm e}}_{4-\ell}$ $(\Delta_n=\tfrac{4\pi}{3}\tfrac{n}{L})$. 
The real part of $\mathcal T^+$ consists of four parts and each of which contributes equally. The data with four different initial configurations with 10$^6$ MCS are shown and the shadows represent their approximate error bars as a guide to eyes. }
\label{fig:ICTQ-Tau_realimag}
\end{figure}



\begin{figure*}[t!]
\includegraphics[width=0.9\textwidth]{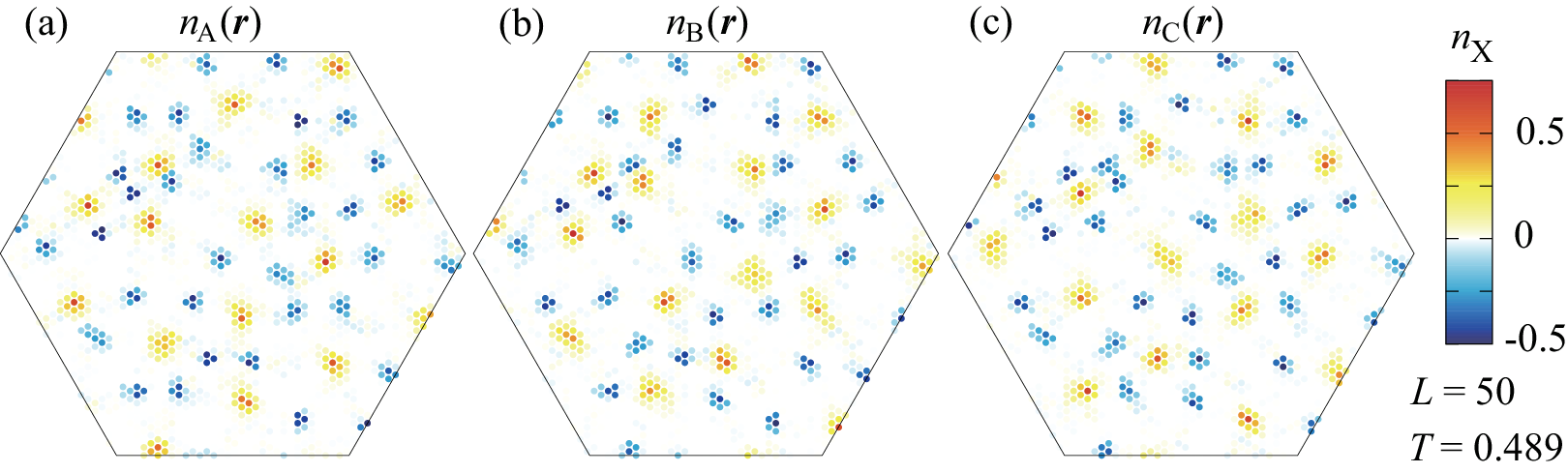}
\caption{Snapshots of sublattice vortex charge configurations at $T=0.489$. The data are averaged over 1000 MCS for $L=50$, $\alpha=0.0938\pi$, and $\lambda=1$.}
\label{fig:vortexlattice}
\end{figure*}



\begin{figure*}[t!]
\includegraphics[width=0.9\textwidth]{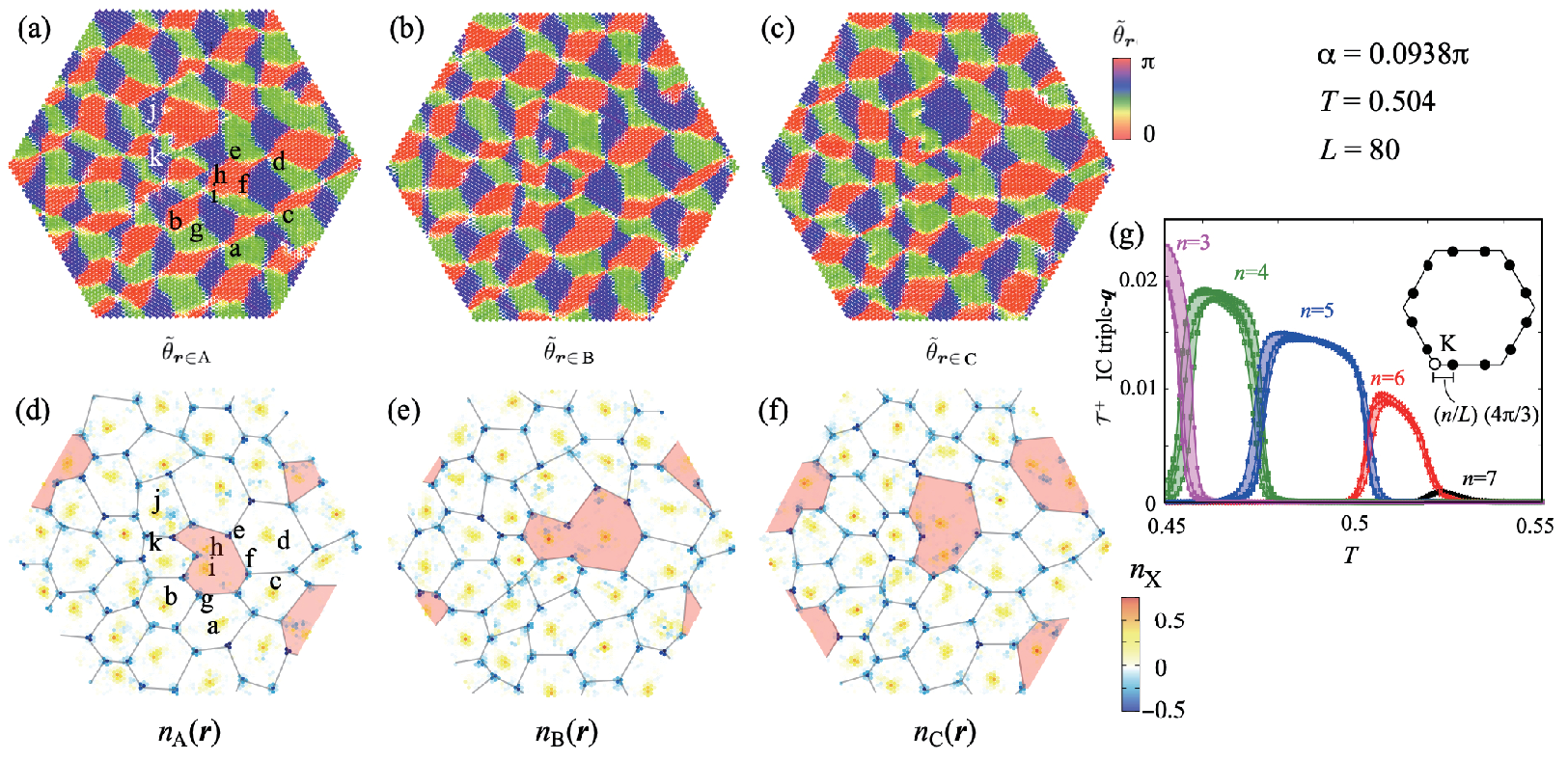}
\caption{
Snapshots of the order parameter angles $\tilde{\theta}_{\rr}$ 
in (a)-(c) and vortex charges $n_{\rm X}(\rr)$ in (d)-(f) 
at $T=0.504$ in the system with 
$\alpha=0.0938\pi$, $L=80$, and $\lambda=1$.
The data are averaged over 1000 MCS.
In (d)-(f), the vortices are connected by lines 
for guide to the eye, and dislocation positions 
are indicated by red shade. 
In (a) and (d), several vortices are labeled by letters a--k. 
(g) Temperature dependence of the 
IC-TQ structure factor $\mathcal{T}^+$.  
Except $T$, all the parameters are identical to those 
for (a)-(f).
These are calculated 
for four ensembles annealed for 10$^6$ MCS 
starting from different initial configurations.   
 Their statistical errors
are shown by shadows as guide to the eye. Inset of (g): the position of the IC ordering vector is shown by filled circles in the 1BZ.
The data
in (a)-(f) corresponds to the boundary between 
the $n=5$ and 6 parts.  
}
\label{fig:dislocation}
\end{figure*}


First, let us analyze the detail of the triple-$\qq$ 
structure factor $\mathcal T^+(\{\qq^\star_\ell\})$. 
As we have explained in the main text, it contains four 
terms as classified by the real ($\phi'_+$) and 
imaginary ($\phi''_+$) parts of the field $\phi_+(\qq_\ell)$ 
\begin{align}
&\mathcal{T}^+(\{\qq^\star_\ell\}) 
= 
\langle 
\phi'_+ (\qq _1^\star) 
\phi'_+ (\qq _2^\star) 
\phi'_+ (\qq _3^\star) 
\rangle
\nonumber\\
&\hspace{1cm}
- \bigl[ 
\langle 
\phi' _+ (\qq _1^\star)
\phi''_+ (\qq _2^\star) 
\phi''_+ (\qq _3^\star) 
\rangle +
(\mbox{cyclic in $\qq^\star_\ell$}) \, \bigr].
\label{eq:TauPlus}
\end{align}
Here, $\langle \cdot \rangle$ denotes the MC average. 
Figure \ref{fig:ICTQ-Tau_realimag} shows $T$-dependence of 
these four terms. The colors [magenta ($n=2$), green ($n=3$), blue ($n=4$)] indicate the structure factor with different ordering vectors $\qq^\star=\qq_{{\rm K}^\prime}-\Delta_n \hat{{\bm e}}_{4-\ell}$ with $\Delta_n=\tfrac{4\pi}{3}\tfrac{n}{L}$. One can see the ordering vector gradually approaches the K$^\prime$(K) point as $T$ decreases. Note that 
the data manifest the relation that the four terms in Eq.~(\ref{eq:TauPlus}) have an identical value, 
$\langle \phi'_+(\qq_1^\star)\phi'_+(\qq_2^\star)\phi'_+(\qq_3^\star)\rangle$$=$$-\langle \phi'_+(\qq_\ell^\star)\phi''_+(\qq_{\ell'}^\star)\phi''_+(\qq_{\ell''}^\star)\rangle$, irrespective of the commensurability $n$. We can understand 
this relation by assuming that the phases $\eta_\ell$ in 
the IC triple-$\qq$ states are not locked but fluctuate over 
all the possible values. When $\phi_\ell$'s are all identical, 
one can set 
$\phi'_+(\qq^\star_\ell)+i\phi''_+(\qq^\star_\ell)=\phi\ e^{i\eta_\ell}$ 
with $\eta_{\rm tot}=\eta_1+\eta_2+\eta_3=2\pi\times$(integer). 
The constraint of the phase sum $\eta_{\rm tot}$ comes from the energy minimization in Eq.~(\ref{eq:FICtQ}). 
This leads to 
\begin{align}
	\mathcal T^+(\{\qq^\star_\ell\})&\propto\langle \cos\eta_1\cos\eta_2\cos\eta_3\rangle\nonumber\\
	&-\big[\langle \cos\eta_1\sin\eta_2\sin\eta_3\rangle+({\rm cyclic\ in\ }\eta_\ell) \ \big]. \label{eq:TauAv}
\end{align}
It is easy to check that no set of fixed values of $\eta_\ell$  
satisfies the relation observed in our simulation. 
Instead, when integrating $\eta_\ell$'s under the assumption of 
uniform realization of all the possible values, we find 
\begin{subequations} 
\begin{align}
&\!\!\int_{-\pi}^\pi \!\! \frac{d\eta_\ell}{2\pi}	
 \int_{-\pi}^\pi \!\! \frac{d\eta_{\ell'}}{2\pi}
\cos\eta_{\ell}\cos\eta_{\ell'}\cos(-\eta_{\ell}-\eta_{\ell'})=\frac{1}{4}, 
\label{eq:TauAv0}\\
&\!\!\int_{-\pi}^\pi \!\! \frac{d\eta_{\ell}}{2\pi}	
 \int_{-\pi}^\pi \!\! \frac{d\eta_{\ell'}}{2\pi}
\cos\eta_{\ell}\sin\eta_{\ell'}\sin(-\eta_{\ell}-\eta_{\ell'})=-\frac{1}{4}. 
\label{eq:TauAv1}
\end{align}
\end{subequations}
This confirms no phase locking of the order parameters in the IC-TQ states.

Next, we visualize vortex configuration in our MC data. 
The sublattice vortex charge $n_{\rm X}(\rr)$  is defined 
in Eq.~(6) in the main text and should be understood as  
a discrete sum along a closed path. 
We take $\rr$ in the sublattice $ {\rm X}$ and consider 
the path formed by 
its nearest-neighbor X sites, i.e., 
$\rr' \in {\rm X}$: hexagon vertices with their center at $\rr$.

Figure \ref{fig:vortexlattice} shows 
a short-time averaged snapshot of $n_{\rm X}(\rr)$ 
sampled over 1000 MCS. The parameters are the same as before, $\alpha=0.0938\pi$, $L=50$, 
$T=0.489$, and $\lambda=1$. 
This is the 
configuration in the region of $n=3$ in 
Fig.~\ref{fig:ICTQ-Tau_realimag}. 
One can see an irregular honeycomb structure made of blue spots 
 with $n_{\rm X}\sim -1/2$ and 
red spots at the center of each hexagon. 
The charge of the red spots is not quantized,  $n_{\rm X}(\rr)<1$, due to the short-time averaging. 
This confirms the formation of vortex lattice in the snapshot configurations of the IC-TQ state. 
The corresponding $\bm{\phi}(\rr)$ is shown in 
Fig.~3(c) in the main text and thus we do not discuss it here.

It is important to see what happens at the points where the ordering vector $\qq^\star_\ell$ jumps in Fig.~\ref{fig:ICTQ-Tau_realimag}. 
 Figures \ref{fig:dislocation}(a)--(c) show the order parameter angle $\tilde{\theta}_{\rr}$ at $T=0.504$ in a larger system with $L=80$. The other parameters are identical to those used in Fig.~\ref{fig:vortexlattice}. This temperature is the point where $n$ changes from 6 to 5 inside the IC-TQ state. As shown in Fig.~\ref{fig:dislocation}(g), $n$ varies as $T$ decreases. See similar behavior  for $L=50$ in Fig.~\ref{fig:ICTQ-Tau_realimag}. In Fig.~\ref{fig:dislocation}(a), some vortex cores are labeled by letters. Let us focus on the islands with  green color. They align along a set of lines running slightly up diagonally to the left. One can see a dislocation at the position h, and a new line of green islands starts from there. A dislocation also exists in the panels (b) and (c) at the same position in the vortex lattices.


\begin{figure*}[t!]
\includegraphics[width=0.8\textwidth]{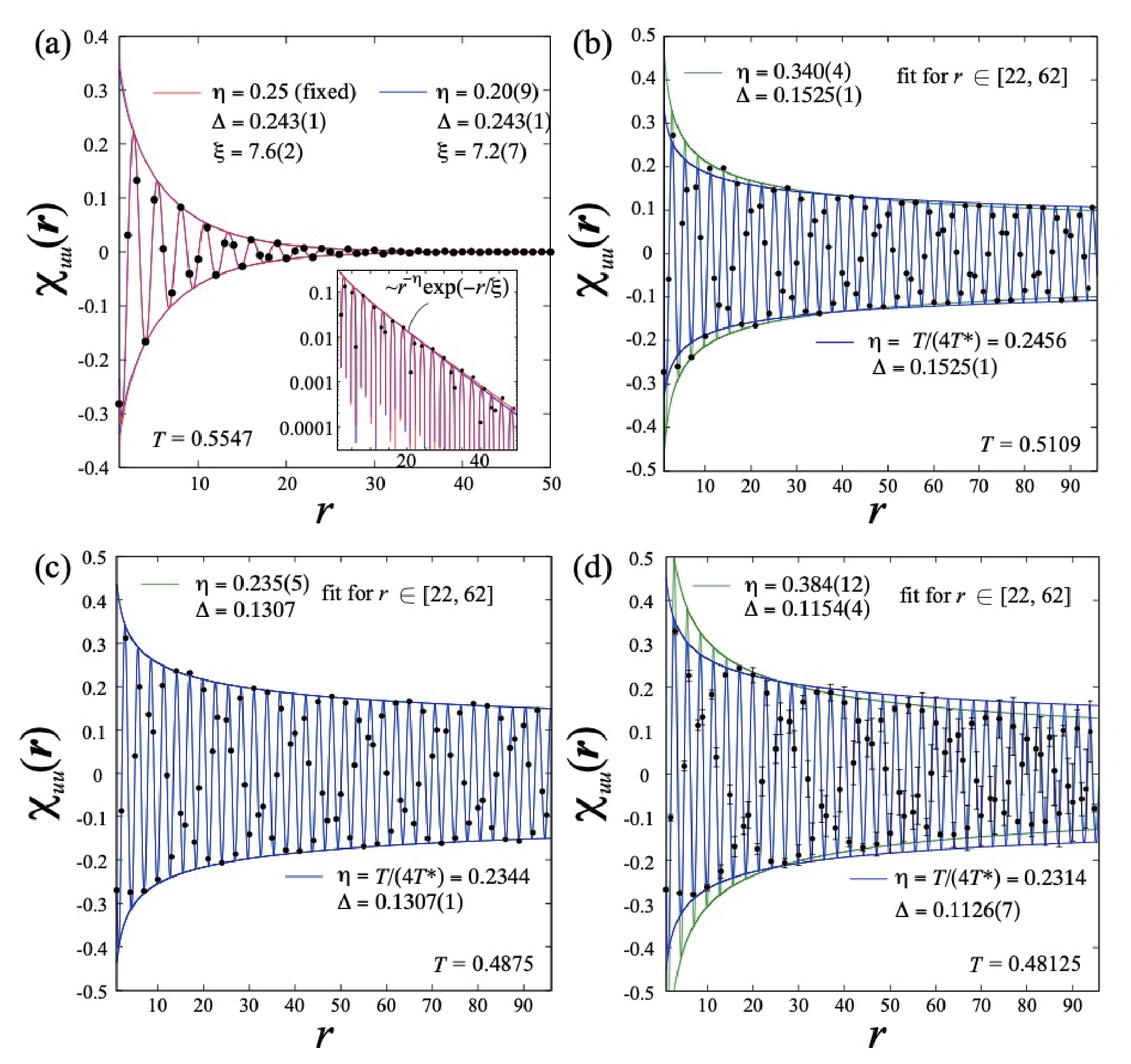}
\caption{
Real-space correlation function $\chi_{uu}(\rr)$ for
$\rr=r \hat{\bm e}_0$ in the system with $L=96$. 
The data for $\rr$ along  the 
equivalent directions 
are averaged to reduce  statistical errors. 
 $T$ is (a) 0.5547, (b) 0.5109, (c) 0.4875, and (d) 0.48125, 
while $T^*\simeq 0.518$. 
Filled circles are the Monte Carlo data with averaging 
over eight sets of 3$\times 10^6$ MCS. 
Lines in the panels are 
$\sim e^{-r/\xi} \cos[(\Delta/2+2\pi/3)r]r^{-\eta}$ in (a) 
and $\sim \cos[(\Delta/2+2\pi/3)r]r^{-\eta}$ in (b)-(d), 
and the parameter values are shown in each panel. 
Inset in (a) is a semilogarithmic plot.  }
\label{fig:IC-sq}
\end{figure*}


 To visualize dislocations more clearly, we have calculated the vortex charge distribution 
 and show the results  in Figs.~\ref{fig:dislocation}(d)--(f). Blue spots are vortex cores and we have connected them by gray lines to visualize a honeycomb-like structure of their arrangement. The non-hexagon areas in light red are the regions including dislocations.  It is well-known that such dislocation formation in two-dimensions indeed destroys long-range order of crystal \cite{Nelson1979-af}. Thus, a two-dimensional ``crystal'' acquires only a quasi-long-range order via the Kosterlitz-Thouless (KT) transition at $T=T^*$ \cite{KT1973}. In the theory of the commensurate-IC transition, it was predicted that even a KT transition does not occur and the system remains a liquid in some situations \cite{Coppersmith1982-im}. In our MC simulations, it is hard to check whether this phase is critical or not by examining the vortex degrees of freedom such as the vortex charge correlation function. This is because the maximum system size $L\sim 100$ is still too small in the sense that the number of vortices is not sufficient to study whether they form a crystal or remain disordered.

 In contrast, orbital degrees of freedom $\bm{\phi}$ exhibit a trace of quasi-long-range order. We consider their fluctuations in the thermodynamic limit are not qualitatively influenced by the variation of IC ordering vector $\qq^\star$. The vector $\qq^\star$ is expected to change smoothly without discrete jumps and the correlation functions of $\bm{\phi}$ exhibit quasi-long-range behavior similar to the KT phase in the XY  model \cite{KT1973}. However, it is not easy to analyze 
 the exponent in the correlation functions, since the change of  $\qq^\star$ remains discontinuous  with lowering $T$ in finite size systems. This makes it difficult to find the behavior in the thermodynamic limit based on finite-size simulations. Nevertheless, as will be explained below, our results show the existence of the KT behavior in the orbital $\bm{\phi}$ sector.

 Let us now discuss the real-space correlation function in the IC-TQ state for $K>0$,
 \begin{align}
 	\chi_{ij}(\rr)\equiv \langle \phi_{i,\rr}\phi_{j,\bm{0}}\rangle,\quad (i,j\in \{u,v\}).
 \end{align}
 To simplify the following discussion, we concentrate on $\chi_{uu}(\rr)$ for $\rr =(r,0)$ with $r$ being an integer.  

 The actual Monte Carlo data of $\chi_{uu}(\rr)$ are well fitted by power-law dependence on $r$ with an oscillation factor as
\begin{align}
	\chi_{uu}(\rr)\propto \frac{1}{r^\eta}\cos\left[\left(\frac{2\pi}{3}+ \frac{\Delta}{2}\right)r\right]\quad {\rm for}\quad T<T^*.\label{eq:chi11}
\end{align} 
For $T>T^*$, 
  Eq.~(\ref{eq:chi11}) is easily extended to that  by including the multiplication factor $\exp(-r/\xi)$ with $\xi$ being the correlation length.

\begin{figure}[t!h]
\includegraphics[width=0.45\textwidth]{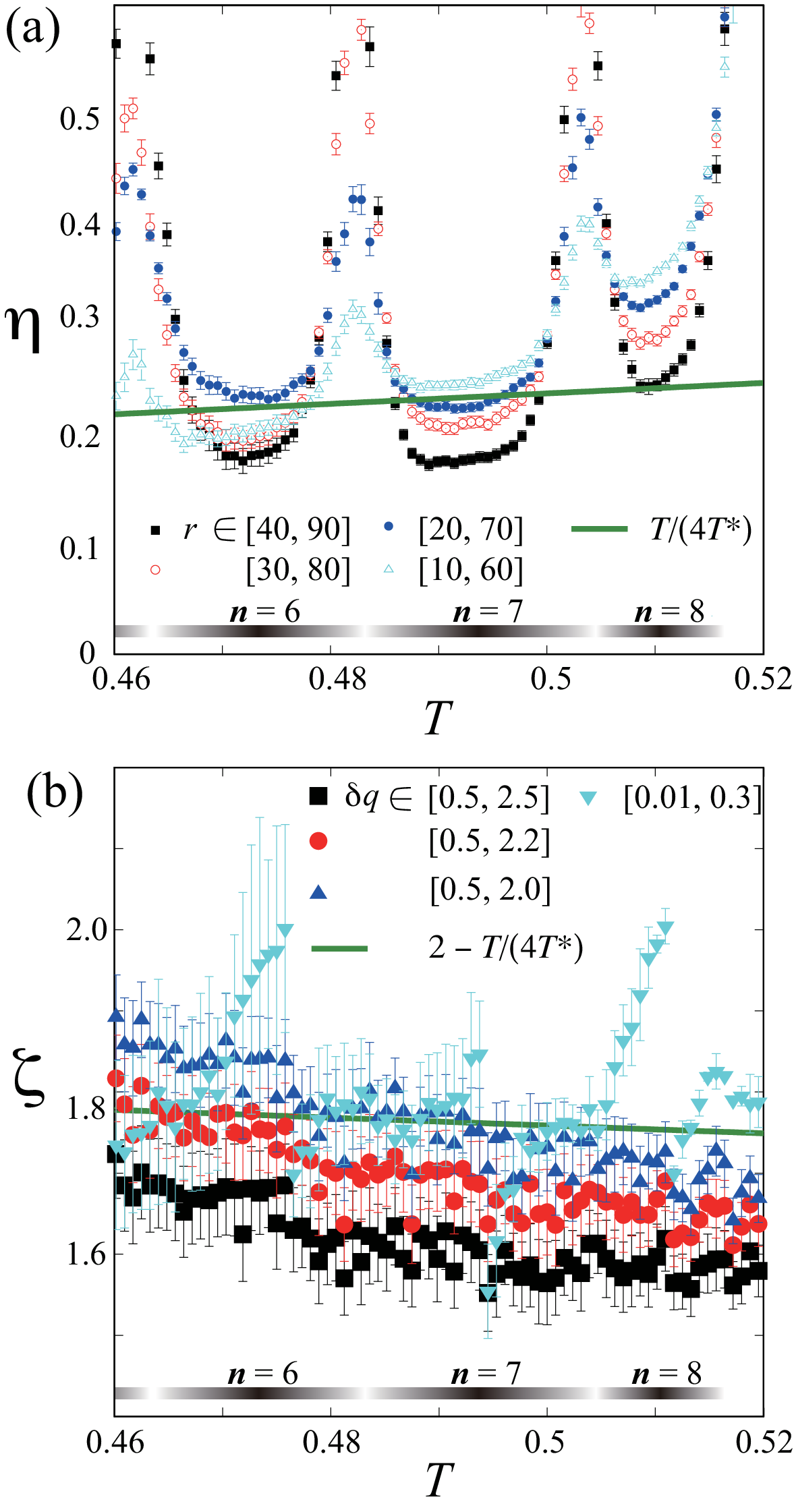}
\caption{
Temperature dependence of estimated exponents $\eta$ and $\zeta$ 
for the system with $L=96$. 
 They define critical behaviors of the correlation functions as 
$\chi_{uu}(\rr) \sim r^{-\eta}$ and 
$\mathcal{S}^+(\qq)\sim |\delta \qq|^{\zeta}$ 
below $T^*=0.518$. 
 The four sets correspond to different fitting ranges $r \in [r_0,r_1]$ in (a) and 
$\delta q\equiv |\qq-\qq^\star|\in [q_0,q_1]$ in (b). 
Green lines indicate 
$\eta _{\rm{KT}} =\frac{1}{4}(T/T^*)$ and 
$\zeta _{\rm{KT}} = 2-\frac{1}{4}(T/T^*)$ 
expected for the KT phase. 
The parameter $n$ is a parameter for $\Delta_{n}$ 
and specifies the $\qq^\star$ vector position. 
The peak structure near the boundaries between two 
 regions with different $n$'s is due to the presence of 
two different modulations in the system.}
\label{fig:IC-exponent}
\end{figure}

  Figure \ref{fig:IC-sq} shows the Monte Carlo data of $\chi_{uu}(\rr)$, and the result of their fitting with Eq.~(\ref{eq:chi11}). 
  One should note that the panel (d) shows the result for 
  the parameter set near the phase boundary to another phase with different $\qq^\star$.  Except for this, the fitting works well with the expression (\ref{eq:chi11}) in (b) and (c), and with the corresponding expression for $T>T^*$ in (a). For comparison, the fitting with $\eta_{\rm KT}=T/(4T^*)$ ($T^*\simeq 0.518$), expected in the simple scaling analysis of the KT transition, is also shown with green curves in (b)--(d) for $T<T^*$ together with those obtained via fitting for $22\le r \le 62$. In (c), the $\eta$ value is almost identical to the expected value $\eta_{\rm KT}$, while larger in (b) in the region of $n=8$ [Fig.~\ref{fig:IC-exponent}(a)]. The latter is due to the fact that the phase with $n=8$ is narrow and the structure factors $\mathcal S^+(\qq^\star_\ell)$ and ${\mathcal T}^+(\{\qq^\star_\ell\})$ are much smaller than those in the other low-$T$ phases. Nevertheless, the difference between the green and blue curves is rather small  in (b) for the intermediate distance  regime $30 \lesssim r \lesssim 70$, where the short-range correlation and the effects of the periodic boundary are both small. For lower $T$, whether the simple form $\eta_{\rm KT}(T)$ is valid or not is nontrivial, and the quantitative discussion well below $T^*$ needs more elaborate and careful analysis. Recently, such analysis has been done in a frustrated transverse-field Ising model\cite{PhysRevB.109.L140408}, where their numerical data show $\eta\sim \eta_{\rm KT}$ even for low $T$ regime in the expected KT phase.


\begin{figure*}[t!]
\includegraphics[width=0.8\textwidth]{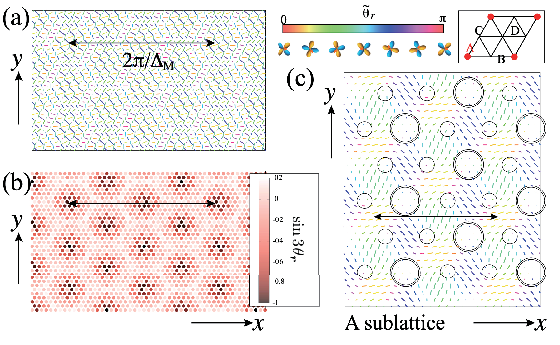}
\caption{
Several configurations in the triple-$\qq$ state with $\qq _{\ell}^\star = 
        \bm{Q}_{\ell} - \Delta_{\rm M} \hat{\bm{e}}_{4-\ell}$ 
with $\Delta_{\rm M} = \frac{\pi}{8}$. Direction of each local orbital/director 
is shown by color in the chart at the right top.  
Sublattice labels are also shown there. 
(a) The director angle $\tilde{\theta}_{\rr}$.  
The length of each bar represents $|\bm{\phi}(\rr)|$.   
(b) The color plot of the potential $\sin3\theta_{\rr}$.
(c) The A-sublattice structure of (a). 
Vortices and half-vortices are shown by double-line big circles 
and single-line small circles, respectively. 
The periodicity of the configuration is 
shown by a double-headed arrow in each panel. 
}
\label{fig:foursubvortex}
\end{figure*}


  Figure \ref{fig:IC-exponent}(a) shows $\eta(T)$  determined by fitting the Monte Carlo data for the largest system $L=96$. In the fitting, several choices of fitting region have been examined. As we have mentioned, the results show noticeable  deviation from $\eta_{\rm KT}(T)$ (green line) around $T^*\simeq  0.518$ and near the  boundaries between phases with different $\qq^\star$'s. For the temperatures well inside each phase, the data sets plotted in blue are close to $\eta_{\rm KT}$, and they are obtained by fitting for $20\le r\le 70$ inside the $n=6$ and 7 phases. Even above $T^*$, the data can be fitted by Eq.~(\ref{eq:chi11}) if $\xi > L$, but this is an artifact of finite system size. For lower $T$, we expect that the $\qq^\star$ vector further moves toward the K point, but we could not carry out efficient numerical simulations at the lower temperatures.

 The correlation function of $\bm{\phi}(\qq)$ also exhibits an evidence of the quasi-long-range order. Figure \ref{fig:IC-exponent}(b) shows the power-law $T$ dependence of the exponent $\zeta$ for $L=96$ in ${\mathcal S}^+(\qq)\simeq |\delta \qq|^{-\zeta}$ with $\delta \qq\equiv \qq-\qq^\star_\ell$. The exponent $\zeta$ is obtained for several fitting ranges and it corresponds to  $2-\eta$. To reduce the effects of subleading hexagonal anisotropy, the two data sets along $\delta \qq \parallel \hat{\bm e}(\ell \omega)$ and $ \parallel\hat{\bm e}((\ell - \frac{1}{4}) \omega)$ are averaged. 
 ${\mathcal S}^+(\qq)$ has six peaks in the 1BZ. 
 Its dependence of a specific $\qq-\qq^\star_\ell$ 
 is affected by the other peaks at $\qq^\star_{\ell'\ne \ell}$. 
Since this is noticeable at $|\delta \qq|\sim 0.4$, we choose the fitting ranges $0.5\le|\delta \qq|\le 2.5$ and $0.01\le |\delta \qq|\le 0.3$.
 We observe that the data obtained by the shorter-range fit in $\delta \qq$ follow $\eta_{\rm KT}$ dependence (sky-blue triangle), and that all the four fitting schemes show increasing $\zeta$ with lowering $T$, when ignoring the behavior near the phase boundaries between the different $n$'s.

Overall, the results shown in Fig.~\ref{fig:IC-exponent} are qualitatively consistent with the scenario that the KT transition occurs and the IC-TQ state possesses a quasi-long-range order. To quantitatively estimate the exponents and carry out calculations at 
lower temperatures, more elaborate numerical simulations are necessary.

To close this section, let us show a typical moir\'e texture and choose $\qq_{\ell}^\star \simeq
 \bm{Q}_{\ell} - \tfrac{\pi}{8}\hat{\bm{e}}_{4-\ell}$. Figure \ref{fig:foursubvortex}(a) shows $\phi(\rr)$ calculated by the superposition of the three plane waves, 
 $\bm{\phi}_{\rr}= \sum_{\ell} 
\cos(\qq _{\ell} \cdot \rr) 
\bm{v}_{\ell}^+$ with $\eta_\ell=0~(\ell=1,2,3)$, 
 which is not the results of MC simulations. The cubic potential energy $\sin 3\theta_{\rr}$ is also shown in Fig.~\ref{fig:foursubvortex}(b), which exhibits a ``charge'' moir\'e pattern. Note also that $\sin3\theta_{\rr}<0$ at the most of the sites. Similarly to the case discussed in the main text, one defines the vortex charge $n_{\rm X}(\rr)$ for the four sublattices (X=A,B,C,D) and the result for ${\rm X}={\rm A}$ is shown in Fig.~\ref{fig:foursubvortex}(c). 
 This  texture is similar to that in Fig.2(b) in the main text. 
 In our simulation, this type of moir\'e pattern is indeed fragile and only observed in a narrow regime between the sTQ and asTQ states. This might be related to the strong stability of the sTQ state. Thus, for stabilizing this type of moir\'e pattern,
 an alternative model is necessary.

\begin{figure}[t!]
\includegraphics[width=0.45\textwidth]{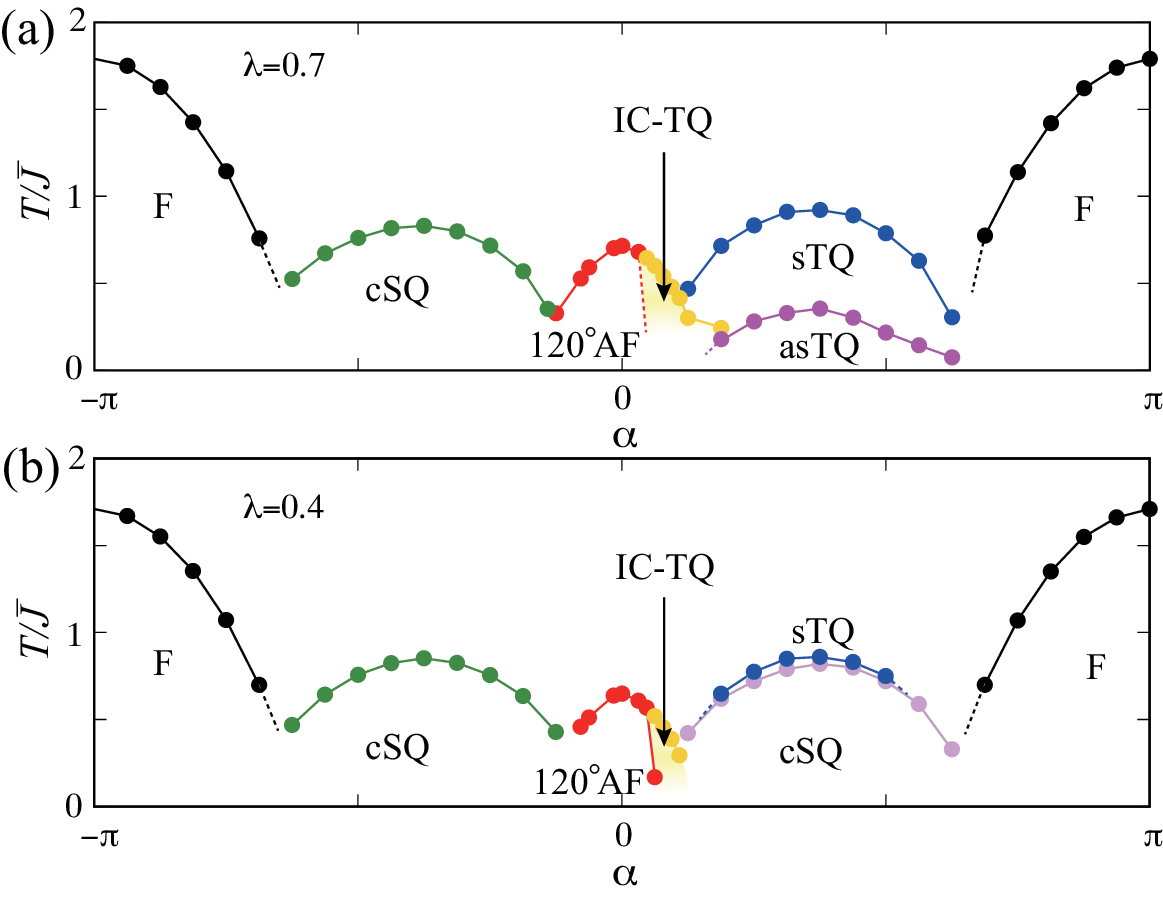}

\caption{
{
$T$--$\alpha$ phase diagram for (a) $\lambda=0.7$ and (b) $\lambda=0.4$. Dashed lines are guide for eyes. Ordered states are ferroic (F), 120$^\circ$ antiferroic (120$^\circ$ AF), canted single-$\qq$ (cSQ), symmetric triple-$\qq$ (sTQ), and asymmetric triple-$\qq$ (asTQ). The two cSQ states in (b) are different each other in their eigenvectors: $\bm{v}^+_{\ell}$'s for $K>0$, while $\bm{v}^-_{\ell}$'s 
for $K<0$.  
}
}
\label{fig:phasediagram2}
\end{figure}

{
\subsection{$\lambda$ dependence of phase diagram}
In the maintext, the cubic coupling $\lambda$ is fixed to $\lambda=1$. Here, we show the phase diagrams in Fig.~\ref{fig:phasediagram2} for smaller values $\lambda=0.7$ and 0.4 to show the overall features do not qualitatively change. Each transition point is determined by the scaling analysis of the structure factor for second order transitions or a jump of the structure factor for first order transitions. We usually use the system size up to $L=48$ and $L=64$ if necessary, for determining the transition temperature of IC-TQ state.

We note that the overall features do not qualitatively change from those for $\lambda=1$. Some  detail differences are as follows. 
(i) The region of the sTQ state for $K > 0$ shrinks toward the higher-$T$ region as $\lambda$ decreases.
This is because the sTQ state is stabilized by the cubic term $\lambda$ and the presence of disordered sites is disfavored at low temperatures for smaller $\lambda$. 
(ii) A single-$\qq$ state with a ferro moment (canted single-$\qq$: cSQ) appears instead of the asTQ for small $\lambda$ and at low $T$. Since the asTQ is also one of triple-$\qq$  states, it is destabilized for smaller $\lambda$. 

 The sTQ and asTQ states are destabilized as $\lambda$ decreases, while the effect on the IC-TQ state is not large at least for $\lambda\ge 0.4$ as shown in Fig.~\ref{fig:phasediagram2}. Exception is the IC-TQ state that exists at the lower temperature side of the sTQ for $\lambda=1.0$. This is replaced by the cSQ for $\lambda=0.4$. However, this IC-TQ state  seems to be stabilized by the detailed free energy competition against the two phases sTQ and asTQ, and we will not discuss here. The IC-TQ order is robust upon decreasing $\lambda$, because there are no competing stable states with the same ordering vector for the intermediate value of $K/J$. This situation clearly differs from that for the sTQ state. There are two such competing states: asTQ and cSQ. Thus, the IC-TQ state is robust even for moderate $\lambda$ 
 and this opens a wider way to finding the orbital moir\'e in material search in future. 
}


\section{D$_6$ point group}
In the last part in the main text, we pointed out the possible patterns of induced polarization and current induced magnetization. Here, we sketch their group theoretical analysis when the local symmetry is the point group D$_6$.   
Let us summarize the irreducible representations made 
from the two-component quadrupole moments $\bm{\phi}=(\phi_u,\phi_v)$, the magnetic moments $\bm{M}=(M_x,M_y,M_z)=:(\bm{M}_\perp,M_z)$, 
and the electric currents $\bm{j}=(j_x,j_y,j_z)=:(\bm{j}_\perp,j_z)$. The character table of the D$_6$ group is shown in Table~\ref{tbl-I}. It is noted that 
the basis sets of the E$_1$ irrep are 
the planer magnetic moment ${\bm M}_\perp$, position (or polarization) $(x,y)$, $\nabla_{\perp}\equiv (\partial_x,\partial_y)$, and planer current $\bm{j}_{\perp}$,  while $\bm{\phi}$ is the E$_2$ irrep. The remaining 
 $j_z$, $M_z$, and $z$ belong to the A$_2$ irrep.

We list below the decomposition of the direct products including $\bm{j}$. 
\begin{itemize}
	\item A$_1$: $j_zM_z,\quad  \bm{j}_\perp \cdot{\bm{M}}_\perp$,
	\item A$_2$: $j_xM_y-j_yM_x=:\bm{j}_\perp \times {\bm{M}}_\perp$,
	\item E$_1$: $\bm{j}_\perp M_z,\quad  j_z{\bm M}_\perp$,
	\item E$_2$: $(j_xM_y+j_yM_x,j_xM_x-j_yM_y)$.
\end{itemize}
For the terms including $\nabla \otimes \bm{\phi}=$E$_1\otimes$ E$_2\oplus$A$_2\otimes $E$_2$, 
\begin{itemize}
	\item B$_1$: $\nabla_\perp \cdot \bm{\phi}  =\partial_x\phi_u+\partial_y\phi_v$,
	\item B$_2$: $\partial_x\phi_v-\partial_y\phi_u=:\nabla_\perp \times \bm{\phi}$,
	\item E$_1$: $(\partial_x\phi_v+\partial_y\phi_u,\partial_x\phi_u-\partial_y\phi_v)=: \bm{\pi}_\perp=(\pi_x,\pi_y)$,
	\item E$_2$: $\partial_z\bm{\phi} =(\partial_z\phi_u,\partial_z\phi_v)$.
\end{itemize}

Now, we consider the quadrupole pattern that transforms in the same way as polarization. That is $\bm{\pi}_\perp$ belonging to the E$_1$ irrep. Its 
contribution to energy under finite $\bm{j}$ is nothing but the invariants including both $\bm{\pi}_\perp$ and $\bm{j}$. They are easily constructed by multiplying its E$_1$ counterpart including $\bm{j}$. 
There are two invariants: $M_z(\bm{\pi}_\perp\times {\bm j}_\perp)$ and  $({\bm M}_{\perp}\times {\bm \pi}_\perp)j_z$. Introducing the three component representation ${\bm \pi}\equiv ({\bm \pi}_\perp,0)$, they read $M_z(\bm{\pi}\times {\bm j})_z$ and  $({\bm M}\times {\bm \pi})_zj_z$ as discussed in the main text.

\begin{table}
\caption{The character table of the D$_6$ point group.}
\label{tbl-I}
\begin{tabular}{ccrrrrrc}
\hline
{\Large \ \ } irrep & 
$E$ & $2 C_6 $ & $2 C_3 $ & $C_2 $ & $3 C_{2 y} $ & $3 C_{2 x} $ & 
basis
\\
\hline
$\mathrm{A}_1,\ \Gamma_1$ & 1 & 1 \ & 1 \ & 1 \ & 1 \ & 1 \ &  $z^2$ 
\\[4pt]
$\mathrm{A}_2,\ \Gamma_2$ & 1 & 1 \ & 1 \ & 1 \ & $-1$ \ & $-1$ \ & $z,M_z$
\\[4pt]
$\mathrm{B}_1,\ \Gamma_3$ & 1 & $-1$ \ & 1 \ & $-1$ \ & 1 \ & $-1$ \ & $y^3-$ $3 x^2 y$ 
\\[4pt]
$\mathrm{B}_2,\ \Gamma_4$ & 1 & $-1$ \ & 1 \ & $-1$ \ & $-1$ \ & 1 \ & $x^3-$ $3 x y^2$ 
\\[4pt]
$\mathrm{E}_1,\ \Gamma_6$ & 2 & 1 \ & $-1$ \ & $-2$ \ & 0 \ & 0 \ & $\{x,y\},\{M_x,M_y\}$ 
\\[4pt]
$\mathrm{E}_2,\ \Gamma_5$ & 2 & $-1$ \ & $-1$ \ & 2 \ & 0 \ & 0 \ & $\{2xy,x^2-y^2\}$ 
\\[4pt]
\hline
\end{tabular}
\end{table}

%
%

\section{Commensurate ``locking''}\label{sec:CL}

As we have discussed in the main text, the IC-TQ state appears for $K>0$ between the AF 120$^\circ$ and the sTQ states. The ordering vectors $\qq^\star$ vary as functions of $T$ and also the interaction parameter $\alpha$. One of the origins of these variations is that the minimum of the exchange coupling $J_{\qq^\star}^-$ is shallow as shown in Fig.~\ref{fig:1}(b). Thus, the energy gain of the quadratic term in the free energy is small and the fourth-order terms become important. 
In Eqs.~(\ref{Fsl-M}), (\ref{Fsl-IC2}), (\ref{eq:FsTQ}), (\ref{eq:FICtQ}), 
(\ref{eq:F120}), and (\ref{eq:FFQ}), one can see the forth-order terms prefer commensurate orders, which manifests commensurate locking. Here, we summarize a simple argument about the commensurate locking 
with taking account of higher order terms in the Landau free energy. 
Consider a real scalar field $\psi(x)$ defined in the one-dimensional 
lattice with length $L$ (lattice constant 1) under 
the periodic boundary condition. In this case, $q$ is quantized as 
$q=2\pi n/L$ with  $n\in\{-L/2+1,\cdots ,0,1,\cdots L/2\}$.

In order to understand the difference between the ``general'' $q$ 
points and the ``high-symmetric'' $q$ points, we consider local 
energy $E_2$ and $E_4$: 
\begin{align}
	E_2=\sum_{j=0}^{L-1}  \psi^2(x_j),\quad {\rm and }\quad 
	E_4=\sum_{j=0}^{L-1} \psi^4(x_j).
\end{align}
Let us consider a simple configuration $\psi(x_j)=\cos(q j)$ 
and examine the corresponding energies. 
We obtain
\begin{subequations}
\begin{align}
\!\!\!\!
E_2 &
=\frac{1}{2} \sum_{j} \bigl[ 1+ \cos(2qj) \bigr] 
=
\begin{cases}
L \quad &\mbox{for $q=0$ and $\pi$}
\\[2mm]
\FRAC12 L \quad &\mbox{for the other $q$'s},
\end{cases}  
\label{eq:E2}
\\
\!\!\!\!
E_4 &
=
\frac{1}{8} \sum_{j} \bigl[ 3+4\cos(2qj)+\cos(4qj) \bigr]
\nonumber\\
\!\!\!\!
&=
\begin{cases}
 L \quad &\mbox{for $q=0$ and $\pi$}
\\[2mm]
\FRAC12 L \quad &\mbox{for  $q=\pm \pi /2$}
\\[2mm]
\FRAC38 L \quad &\mbox{for the other $q$'s}.
\end{cases}
\label{eq:E4}
\end{align}
\end{subequations}
From these expressions, one can find a clear contrast between the general $q$ points and the high-symmetry $q$ points.

Alternatively, we can derive the same results by calculations in 
 $q$-space sum. Let $c_q$ denote the Fourier component of $\psi(x)$. 
 The reality imposes the relation $c_{-q}=c^*_{q}$. First, $E_2$ reads
\begin{align}
	E_2&= \sum_{q}|c_q|^2
		=c_0^2+c_{\pi}^2+ 2\sum_{0<q<\pi}|c_{q}|^2.
\end{align}
Note that $c_0$ and $c_\pi$ are both real and the general-$q$ terms appear twice in the form of $c_{q}c_{-q}$ and  $c_{-q}c_{q}$, while $q=0$ and $\pi$ appear once. These results completely agree with Eq.~(\ref{eq:E2}) by substituting $c_0=\sqrt{L}$, $c_\pi=\sqrt{L}$,  or $c_{q}=\sqrt{L}/2$ for $0<q<\pi$. Next consider $E_4$,
\begin{align}
	\!\!\!\!\!\! E_4  
	&=\frac{1}{L}\sum_{qq'pp'}\sum_G c_qc_{q'}c_{p}c_{p'}\delta_{q+q'+p+p',G}\nonumber\\
		&=\frac{1}{L}\Big[c_0^4+c_{\pi}^4+6{\rm Re}(c_{\pi/2}^4)+6\sum_{0<q<\pi}|c_{q}|^4+\cdots\Big], 
\end{align}
where $\{G=2\pi\times ({\rm integer})  \}$ are the reciprocal lattice vectors. 
Note ${\rm Re}(c_{\pi/2}^4)={\rm Re}(c_{-\pi/2}^4)$, and 
we have written down only terms relevant for the discussion about the  
single-$q$ orders. The factor $6$ arises from the combinatorial number 
choosing two $c_q$'s out of four. $c^4_{\pi/2}$ and $c^4_\pi$ terms arise 
from the Umklapp conditions with $G=\pm 2\pi$ and $G=\pm 4\pi$, respectively. Again, substituting $c_{0}=c_{\pi}=\sqrt{L}$ and $c_{\pm q}=\sqrt{L}/2$ for $0<q<\pi$, the results reproduce Eq.~(\ref{eq:E4}). 
These results demonstrate that the free energy has different coefficients 
 between the general-$q$ and high-symmetry-$q$, and this reflects the tendency of ``locking'' of the ordering vectors to the rational (high-symmetry) ones.

\providecommand{\noopsort}[1]{}\providecommand{\singleletter}[1]{#1}%

\end{document}